\documentclass{elsart}

\usepackage{amssymb,graphicx}
\usepackage{ifthen}
\usepackage{color}

\newboolean{RevTeX}
\setboolean{RevTeX}{false}

\def\p{\partial}
\newcommand{\bm}[1]{\mbox{\boldmath$#1$}}               
\renewcommand{\Re}{{\rm Re}}

\newcommand{\ud}{\, \mathrm d}

\newcommand{\columntwo}[2]
{
  \left (
  \begin{array}{c}
    {#1} \\
    {#2}
  \end{array}
  \right )
}

\newcommand{\columnthree}[3]
{
  \left (
  \begin{array}{c}
    {#1} \\
    {#2} \\
    {#3}
  \end{array}
  \right )
}

\newcommand{\matrixtwotwo}[4]
{
  \left (
  \begin{array}{lll}
    {#1} & \quad & {#2} \\
    {#3} & \quad & {#4}
  \end{array}
  \right )
}

\newcommand{\matrixthreethree}[9]
{
  \left (
  \begin{array}{lll}
    {#1}    & {#2}    & {#3} \\
    {#4}    & {#5}    & {#6} \\
    {#7}    & {#8}    & {#9}
  \end{array}
  \right )
}

\newcommand{\uhat}{\hat u}
\newcommand{\vhat}{\hat v} 
\newcommand{\what}{\hat w}
\newcommand{\D}{D^{(1)}}
\newcommand{\DD}{D^{(2)}}
\newcommand{\Dhat}{\hat D^{(1)}}
\newcommand{\DDhat}{\hat D^{(2)}}

\ifthenelse{\not \boolean{RevTeX}}
{
  \journal{Journal of Computational Physics}
}

\date{18 May 2005}

\begin{document}

\ifthenelse{\not \boolean{RevTeX}}
{
  \begin{frontmatter}
}


\title{Numerical stability for finite difference approximations of
Einstein's equations}



\ifthenelse{\boolean{RevTeX}}
{
\author{Gioel Calabrese}
\author{Ian Hinder}
                                                                                
\affiliation{School of Mathematics, University of Southampton,
Southampton, SO17 1BJ, UK}
                                                                                
\author{Sascha Husa}
\affiliation{Max-Planck-Institut fur GravitationsPhysik,
Albert-Einstein-Institut, Am M\"uhlenberg 1, D-14476, Golm, Germany}
}
{
\author[Soton]{G. Calabrese},
\author[Soton]{I. Hinder},
\author[AEI]{S. Husa},

\address[Soton]{School of Mathematics, University of Southampton,
Southampton, SO17 1BJ, UK}
\address[AEI]{Max-Planck-Institut fur GravitationsPhysik,
Albert-Einstein-Institut, Am M\"uhlenberg 1, D-14476, Golm, Germany}
}

\begin{abstract}

We extend the notion of numerical stability of finite difference
approximations to include hyperbolic systems that are first order in
time and second order in space, such as those that appear in Numerical
Relativity and, more generally, in Hamiltonian formulations of field
theories.  By analyzing the symbol of the second order system, we
obtain necessary and sufficient conditions for stability in a discrete
norm containing one-sided difference operators.  We prove stability
for certain toy models and the linearized Nagy-Ortiz-Reula formulation
of Einstein's equations.

We also find that, unlike in the fully first order case, standard
discretizations of some well-posed problems lead to unstable schemes
and that the Courant limits are not always simply related to the
characteristic speeds of the continuum problem.  Finally, we propose
methods for testing stability for second order in space hyperbolic
systems.

\end{abstract} 

\ifthenelse{\boolean{RevTeX}}
{
  \maketitle
}
{

  \end{frontmatter}
}

\section{Introduction}

The Einstein equations consist of a set of ten coupled non linear
second order partial differential equations.  In order to perform
numerical time evolutions the fully second order system is usually
written as a first order in time system.  Such systems can be evolved
directly \cite{SN,BShap}, or a further reduction from second to first
spatial order can be performed (see, for example, \cite{FR,H,AY,KST}).
Whereas the theory of Cauchy problems for fully first order systems of
partial differential equations is understood, in terms of
well-posedness at the continuum and the stability of finite difference
approximations, the theory of second order in space hyperbolic systems
is less well developed.  The recent improvement in the understanding
of second order in space formulations of Einstein's equations at the
continuum \cite{SCPT,NOR,GG1,GG2,BS}, has not been matched by
developments concerning finite difference approximations of such
systems (see, however, \cite{KPY,SKW}).  Given that these systems have
fewer variables, fewer constraints, and typically smaller errors (see
\cite{KPY} and Appendix \ref{Sec:NumProp12}), it is desirable to
better appreciate their properties.  Note that first order in time
hyperbolic systems, which are not necessarily first order in space,
also arise naturally in Hamiltonian formulations of field theories.

The standard notion of stability for fully first order systems based
on the discrete $L_2$ norm is unsuitable for analyzing second order in
space hyperbolic systems.  This can be understood by analogy with the
continuum result for the one dimensional wave equation written in
first order in time and second order in space form: $\p_t \phi(t,x) =
\Pi(t,x)$, $\p_t \Pi(t,x) = \p_x^2 \phi(t,x)$. Consider the family of
solutions $\phi(x,t) = \sin(\omega x) \cos(\omega t)$, $\pi(x,t) = -
\omega \sin(\omega x) \sin(\omega t)$ generated by the initial data
$\phi_0(x) = \sin(\omega x)$, $\pi_0(x) = 0$.  By varying $\omega$ in
the initial data, the $L_2$ norm of the solution at a fixed time $t$,
$\int_0^{2\pi} (|\phi|^2 + |\Pi|^2) dx$, can be made arbitrarily large
with respect to the initial data (whose norm is independent of
$\omega$), thus contradicting well-posedness of the Cauchy problem in
$L_2$ \cite{RM,FG}. The introduction of the new variable,
$X=\p_x\phi$, allows the construction of a first order system, the
Cauchy problem of which is well-posed in $L_2$.  The original second
order problem can then be shown to be well-posed in a norm containing
derivatives, namely $\int_0^{2\pi} ( |\phi|^2 + |\Pi|^2 + |\p_x
\phi|^2 ) dx$, which corresponds to the $L_2$ norm of the first order
reduction.

In this work we consider linear constant coefficient Cauchy problems.
We use the method of lines to separate the time integration from the
spatial discretization.  We show that by reducing the discrete system
to first order in Fourier space, it is possible to determine stability
in physical space with respect to a discrete norm containing one-sided
difference operators. This is done by extending the notion of a
symmetrizer to the discrete case.  We apply these techniques to
problems, starting with the wave equation written as a first order in
time, second order in space system.  We consider both second and
fourth order accurate discretizations.  A similar but more complicated
analysis is done for the Knapp-Walker-Baumgarte (KWB) \cite{KWB} and
Z1 \cite{Z1} formulations of electromagnetism, and the
Nagy-Ortiz-Reula (NOR) \cite{NOR} formulation of Einstein's equations.
We also point out stability issues related to the ADM \cite{ADM} and Z4
\cite{BLPZ1} formulations.

In Sec.~\ref{Sec:Background}, we summarize some relevant material
from the literature.  In Sec.~\ref{Sec:Stability12} we introduce
the concept of a discrete symmetrizer.  
We also illustrate the reduction procedure to first order in Fourier
space, which can be used for obtaining energy estimates at the
continuum.  We introduce the analogous idea for the discrete case, and
discuss convergence.  In Sec.~\ref{Sec:Applications} we apply these
techniques to the systems mentioned above.  We propose methods in
Sec.~\ref{Sec:Testing} for testing stability experimentally both for
linear and non linear systems.  We summarize the main results of this
paper in Sec.~\ref{Sec:Discussion}.  In Appendix
\ref{Sec:Timeintegrators}, we describe the different time integration
methods that we consider, and in Appendix \ref{Sec:NumProp12} we
compare numerical properties of the wave equation written as a first
order system with those of the wave equation written as a first order
in time, second order in space system.  In Appendix
\ref{Sec:ConstrProp} we highlight differences in the constraint
propagation properties between first and second order systems.

\section{Background}
\label{Sec:Background}

Well-posedness, the (local in time) existence of a unique solution
which depends continuously on the problem's data, is a fundamental
requirement for the successful generation of numerical solutions
approximating the solution of a continuum problem.  In this section we
review the notion of well-posedness for linear constant coefficient
Cauchy problems, as well as the concept of stability for finite
difference approximations.  We conclude the section by providing a
simple sufficient condition for stability of first order fully
discrete problems based on the properties of the symbol of the
semi-discrete system, which will be extended to discretizations of
second order in space problems in the next section.

\subsection{Constant coefficient Cauchy problems}
\label{Sbc:constcauchy}

In this work we will be dealing with initial value (or Cauchy)
problems of the form
\begin{eqnarray}
\frac{\p}{\p t}u(t,x) &=& P\left(\frac{\p}{\p x}\right)
u(t,x)\,, \label{Eq:Cauchy1} \\
u(0,x) &=& f(x)\,, \label{Eq:Cauchy2}
\end{eqnarray}
in $d$ spatial dimensions, where $x \in \mathbb{R}^d$, $u =
(u^{(1)},u^{(2)}, \ldots u^{(m)})^T$ and $P$ is a linear, constant
coefficient, differential operator of order $p$.  We consider only the
cases $p=1$ and $p=2$.  Furthermore, we assume that the eigenvalues of
the symbol of the differential operator, $\hat P(i\omega)$, which is
obtained by replacing $\p/\p x_j$ in $P(\p/\p x)$ with $i\omega_j$,
for $j=1,2,\ldots, d$, have real part uniformly bounded from below and
above.  We are thus excluding parabolic systems, but we are allowing
for systems like the wave equation written as a first order in time,
second order in space system.  For simplicity we focus on solutions
that are $2\pi$-periodic in all spatial coordinate directions.  Thus
the initial data, $f(x)$, is chosen so that it satisfies this
property.

We consider the $p=1$ case, leaving the $p=2$ case for the next
section.  Following Definition 4.1.1 in \cite{GKO-Book} we say that
problem (\ref{Eq:Cauchy1})--(\ref{Eq:Cauchy2}) is well-posed with
respect to a norm $\| \cdot \|$ if for every smooth periodic $f$ there
is a unique smooth spatially periodic solution and there are constants
$\alpha$ and $K$, independent of $f$, such that for $t\ge 0$
\begin{equation}
\| u(t,\cdot)\| \le Ke^{\alpha t} \| f\|\,. \label{Eq:estimate}
\end{equation}
Exponential growth must be allowed if one wants to treat problems with
lower order terms.  For first order hyperbolic systems the $L_2$ norm
$\| w \|^2 = \int_0^{2\pi} \dots \int_0^{2\pi} | w(x)|^2 dx_1\ldots
dx_d$ is usually used in (\ref{Eq:estimate}).  We will see later that
the second order systems we study in this work require the use of a
different norm.

Taking $f(x) = (2\pi)^{-d/2} \sum_{\omega} e^{i\langle \omega, x
\rangle} \hat{f}(\omega)$ the formal solution of
(\ref{Eq:Cauchy1})--(\ref{Eq:Cauchy2}) is $u(t,x) = (2\pi)^{-d/2}
\sum_{\omega} e^{i\langle \omega, x \rangle} e^{\hat{P}(i\omega)t}
\hat{f}(\omega)$.  It can be shown (Theorem 4.5.1 in \cite{GKO-Book})
that well-posedness in the $L_2$ norm is equivalent to there being
constants $K$, $\alpha$ such that, for all $\omega$ and for $t \ge 0$,
\begin{equation}
| e^{\hat{P}(i\omega)t}| \le Ke^{\alpha t}, \label{Eq:Kealphat}
\end{equation}
where $|A| = \sup_{|u|=1} | Au |$ is the matrix (operator) norm of a
matrix $A$.

Well-posedness of the Cauchy problem in the $L_2$ norm is also
equivalent (Theorem 4.5.8 in \cite{GKO-Book}) to the existence of
constants $\alpha$, $K>0$ and of Hermitian matrices $\hat H(\omega)$
satisfying\footnote{Two Hermitian matrices, $A$ and $B$, satisfy $A
\le B$ if and only if $x^* A x \le x^* B x$ for every $x$.  If a
matrix $\hat H(\omega)$ satisfies $K^{-1}I \le \hat H(\omega) \le KI$
for every $\omega$, we say that $\hat H(\omega)$ is {\em equivalent to
the identity matrix}.}, for every $\omega$,
\begin{eqnarray}
&&K^{-1}I \le \hat H(\omega)
\le KI\,,\label{eqn:contsym}\\
&&\hat H(\omega) \hat P(i\omega) + \hat
P^*(i\omega) \hat H(\omega) \le 2\alpha \hat H(\omega)\,,
\nonumber
\end{eqnarray}
where $\hat P^*$ represents the Hermitian conjugate of $\hat P$.  The
last inequality gives an energy estimate for each Fourier mode and the
estimate in physical space, Eq.~(\ref{Eq:estimate}), follows from
Parseval's relation, $\| u (t,\cdot)\|^2 = \sum_{\omega} | \hat u
(t,\omega) |^2$.  Since the existence of $\hat H(\omega)$ is not
affected by the addition of a constant matrix to $\hat P(i\omega)$
(Lemma 2.3.5 in \cite{KL-Book}), undifferentiated terms on the right
hand side of the equations can be ignored in the analysis of
well-posedness.  If (\ref{eqn:contsym}) is satisfied with $\hat H \hat
P + \hat P^* \hat H = 0$ then $\hat H$ is called a {\em symmetrizer}.

For $p=1$, system (\ref{Eq:Cauchy1}) is said to be {\em strongly
hyperbolic} if the corresponding Cauchy problem is well-posed in the
$L_2$ norm (i.e.~if $\hat H(\omega)$ exists) \footnote{For $\p_t u =
A^i \p_i u$ the symbol is $\hat P = i \omega_i A^i$.  The system is
said to be weakly hyperbolic if the eigenvalues of $\hat P(i\omega)$
are imaginary.  Strong hyperbolicity is equivalent to $\hat P(i
\omega)$ being uniformly diagonalizable with imaginary eigenvalues.
We define the characteristic speeds in the direction $\omega_i$ to be
the eigenvalues of $\hat P(i\omega)$ divided by $i\omega$.}.  If $\hat
H(\omega) = I$, the system is said to be {\em symmetric hyperbolic}.
If $\hat H(\omega) = H$ is independent of $\omega$, then we say that
the system is {\em symmetrizable hyperbolic}\footnote{Symmetrizable
hyperbolic systems are often also called symmetric hyperbolic.}. In
this case the change of variables $\tilde u = H^{1/2} u$ brings the
system into symmetric hyperbolic form.  Finally, well-posedness is not
affected by the presence of forcing (inhomogeneous) terms (Theorem
4.7.2 in \cite{GKO-Book}).  For cases where such terms are present,
the estimate requires modification.

Note that, in the absence of lower order terms, whereas symmetrizable
hyperbolicity guarantees the existence of a conserved energy in
physical space, $(u,Hu)$, a strongly hyperbolic system satisfies the
estimate $\| u(t,\cdot)\| \le K \| u(0,\cdot)\|$ with a constant $K
\ge 1$.  Furthermore, in the variable coefficient case,
well-posedness results require smoothness of the symmetrizer $\hat
H(x,t,\omega)$ in all arguments \cite{KL-Book}.

\subsection{Numerical stability}

\subsubsection{Notation}

Our notation and conventions follow closely those of \cite{GKO-Book}.
We introduce a spatial grid $x_j = (x^{(1)}_{j_1}, x^{(2)}_{j_2},
\ldots, x^{(d)}_{j_d}) = (j_1h_1,j_2h_2,\ldots,j_dh_d)$, where $h_r =
2\pi/N_r$ and $j_r = 0, 1, \ldots, N_r-1$, and the vector-valued grid
function $v_j(t)$ approximating $u(t,x_j)$.  Periodicity requires
that $v_j = v_{{\rm mod}(j,N)}$.  The partial derivatives in
(\ref{Eq:Cauchy1}) are approximated using either the {\em standard
second order accurate discretization}
\begin{eqnarray}
&&\p_i \to D_{0i}\,,\qquad \p_i\p_j \to
\left\{ \begin{array}{ll}D_{0i}D_{0j} & \mbox{if $i \neq j$}\\
D_{+i}D_{-i} & \mbox{if $i =j$}\end{array}\right.\,,
\label{Eqstandiscr2}
\end{eqnarray}
or the {\em standard fourth order accurate discretization}
\begin{eqnarray}
&& \p_i \to D^{(4)}_i \equiv D_{0i}\left( 1-
  \frac{h^2}{6}D_{+i}D_{-i}\right)\,, \label{Eqstandiscr4}\\ 
&&\p_i\p_j \to \left\{ \begin{array}{ll} D^{(4)}_i D^{(4)}_j& \mbox{if
      $i \neq j$}\\ D_{+i}D_{-i}
  \left(1-\frac{h^2}{12}D_{+i}D_{-i}\right) 
& \mbox{if $i =j$}\end{array}\right.\,,\nonumber
\end{eqnarray}
where $D_+v_j = (v_{j+1}-v_j)/h$, $D_-v_j = (v_{j}-v_{j-1})/h$,
$D_0v_j = (v_{j+1}-v_{j-1})/2h$, and $D_+D_-v_j =
(v_{j+1}-2v_j+v_{j-1})/h^2$. The discretization of $\p_i^2$ as in
(\ref{Eqstandiscr2}) or (\ref{Eqstandiscr4}) gives the desired order
of local accuracy without requiring a larger stencil.  We then
integrate the resulting system of $m\prod_{r=1}^d N_r$ ordinary
differential equations
\begin{eqnarray}
\frac{d}{dt} v_j(t) &=& P v_j(t)\,,\label{Eq:semi1}\\
v_j(0) &=& f_j\,,\label{Eq:semi2}
\end{eqnarray}
where $f_j = f(x_j)$, with three different time integrators.  These
are iterative Crank Nicholson (ICN) and third and fourth order
Runge-Kutta (3RK and 4RK) methods, which are widely used by numerical
relativists (see Appendix \ref{Sec:Timeintegrators} for definitions).
Using the fact that the operator $P$ is linear and time independent we
can write the fully discrete system in polynomial form (see for
example \cite{GKO-Book})
\begin{eqnarray}
v^{n+1}_j &=& Q v^n_j = {\mathcal P} (kP) v^n_j\,, \label{Eq:scheme1}\\
v^0_j &=& f_j\,, \label{Eq:scheme2}
\end{eqnarray}
where $k = \lambda h$ is the time step, $\lambda$ is called the
{\em Courant factor}, and $v^n_j$ represents the grid-function at time $t_n
= nk$.  This is an explicit, one step, scheme.  For ICN we have
${\mathcal P}(x) = 1+ 2\sum_{r=1}^3 \frac{x^r}{2^r}$, whereas for
$p$-th order Runge-Kutta we have ${\mathcal P}(x) = \sum_{r=0}^p
\frac{x^r}{r!}$.

\subsubsection{Definition of stability}

We recall the definition of numerical stability and
discuss some necessary and sufficient conditions. The solution of the
finite difference scheme (\ref{Eq:scheme1})--(\ref{Eq:scheme2}) is
$v^n = Q^n f$.  We introduce the scalar product $(u,v)_h = \sum_j
\langle u_j, v_j \rangle h^d$, where $h^d = \prod_{i=1}^d h_i$,
$j=(j_1,j_2,\ldots, j_d)$ is a multi-index and $\langle u_j,v_j
\rangle = \sum_{r=1}^m \bar{u}_j^{(r)} v_j^{(r)}$.  This allows us to
define a norm $\| v \|_h = (v,v)_h^{1/2}$.  The approximation
(\ref{Eq:scheme1})--(\ref{Eq:scheme2}) is said to be {\em stable} with
respect to this norm if there exist constants $\alpha$, $K$, such that
for all $h$, $k$, $0 < h \le h_0$, $0 < k \le k_0$, the estimate
\begin{equation}
\| v^n \|_h \le K e^{\alpha t_n} \| f \|_h \label{Eq:stability}
\end{equation}
holds for all $n$ such that $t_n = nk$ and all initial
grid-functions $f$.  This concept of stability is the discrete
analogue of (\ref{Eq:estimate}).  It guarantees that the solutions are
bounded as $h\to 0$.  However, the schemes we consider are at most
conditionally stable.  By this we mean that there exists a $\lambda_0$
such that the above inequality holds if and only if the additional
condition $\lambda = k/h \le \lambda_0$ is satisfied.

Theorem 5.1.2 in \cite{GKO-Book} guarantees that if the scheme
(\ref{Eq:scheme1})--(\ref{Eq:scheme2}) is stable, then the modified
scheme
\begin{eqnarray}
v^{n+1}_j &=& (Q + k R) v^n_j\,, \label{Eq:schemelot1}\\
v^0_j &=& f_j \label{Eq:schemelot2}
\end{eqnarray}
is also stable provided that $R$ is bounded.  This will be the case
when $R$ represents constant terms (lower order terms) in the
continuum problem.  Hence for a first order in space system
lower order terms can be ignored without affecting stability.

\subsubsection{Convergence}

Following Theorem 5.1.3 in \cite{GKO-Book}, consistency and stability
imply convergence.  Assume that the continuum solution $u$ of
(\ref{Eq:Cauchy1})--(\ref{Eq:Cauchy2}) is
smooth and that the scheme (\ref{Eq:scheme1})--(\ref{Eq:scheme2}) is
stable.  Further assume that the scheme and the initial data are
consistent. Then, on any finite interval $[0,T]$, the error satisfies
\begin{equation}
\|v^n - u(\cdot, t_n)\|_h \le O(h^{p_1} + k^{p_2}) \label{Eq:error}
\end{equation}
i.e. the solutions of the finite difference scheme converge as $h \to
0$ to the solution of the differential
equation\footnote{Note that the big $O$ in inequality
(\ref{Eq:error}) contains higher derivatives of the exact solution.
Smoothness of the solution of the continuum problem is not required
for convergence.  For instance, a weaker condition for fourth order
convergence ($p_1=p_2=4$) is that the solution be $C^5$.}.

\subsubsection{Fourier analysis of stability}
\label{Sec:FAS}

For approximations with constant coefficients, Fourier analysis can be
used to obtain necessary and sufficient conditions for stability which
can be more easily verified than the above definition.  We assume that
$N$, the number of grid-points in each direction, is even (the odd case
is discussed in Sec.~\ref{Sec:oddcase}).  If we represent $v^n_j$ by
\begin{equation}
v^n_j = \frac{1}{(2\pi)^{\frac{d}{2}}} \sum_{\omega} e^{i\langle\omega,
  x_j\rangle} \hat{v}^n (\omega), \label{Eq:intpol}
\end{equation}
where $\omega = (\omega_1, \omega_2, \ldots, \omega_d)$, $\omega_r =
-N/2+1,\ldots, N/2$, and substitute it into the difference scheme
(\ref{Eq:scheme1})--(\ref{Eq:scheme2}), we obtain
\begin{eqnarray}
\hat{v}^{n+1}(\omega) &=& \hat{Q}(\xi) \hat{v}^n(\omega)\label{Eq:Fscheme1},\\
\hat{v}^0(\omega) &=& \hat{f}(\omega), \label{Eq:Fscheme2}
\end{eqnarray}
where $\xi_r = \omega_r h = -\pi + 2\pi/N,-\pi + 4\pi/N, \ldots, +\pi$
and $r=1,2,\ldots, d$.  The $m\times m$ matrix $\hat{Q}(\xi)$ is
called the {\em amplification matrix} of the scheme and is a real
polynomial in $\hat P$, the symbol of the Fourier transformed
semi-discrete problem,
\begin{equation}
\hat Q(\xi) = {\mathcal P}(k\hat P(\xi))\,. \label{Eq:hatQ}
\end{equation}
The matrix $\hat P(\xi)$ will play an important role in the next
section.  It can be readily computed from $P$ in Eq.~(\ref{Eq:semi1})
with the replacements
\begin{eqnarray}
D_{0i} &\to& \frac{i}{h} \sin \xi_i,\\
D_{+i}D_{-i} &\to& -\frac{4}{h^2} \sin^2 \frac{\xi_i}{2}.
\end{eqnarray}

Using the discrete Parseval's relation
\begin{equation}
\|v\|^2_h = \sum_{\omega} |\hat{v}(\omega)|^2 \label{Eq:discrParseval}
\end{equation}
and the fact that the solution of
(\ref{Eq:Fscheme1})--(\ref{Eq:Fscheme2}) is $\hat{v}^n(\omega) =
\hat{Q}^n(\xi) \hat f(\omega)$ one can show (Theorem 5.2.1 of
\cite{GKO-Book}) that a necessary and sufficient condition for
stability with respect to the $\|\cdot\|_h$ norm is given by
\begin{equation}
| \hat{Q}^n(\xi) | \le K e^{\alpha t_n} \label{Eq:hatQn}
\end{equation}
for all $h = 2\pi/N \le h_0$, $k\le k_0$, $n$ with $t_n=nk$, and
$\omega_r = - N/2+1,\ldots, N/2$, $r=1,2,\ldots, d$.

A much easier condition to verify is the von Neumann condition, which
is only a necessary condition for stability.  It corresponds to the
requirement that the eigenvalues $z_\nu(\xi)$ of $\hat{Q}(\xi)$
satisfy
\begin{equation}
|z_\nu(\xi)| \le e^{\alpha k} \label{Eq:vonNeumann}
\end{equation}
for all $h \le h_0$ and $|\xi_r| \le \pi$.  However, when the
amplification matrix can be uniformly diagonalized (i.e. there exists
a non-singular matrix $T(\xi)$ that diagonalizes $\hat Q(\xi)$ and
satisfies $|T(\xi)||T^{-1}(\xi)| \le C$ with $C$ independent of $\xi$)
then the von Neumann condition is also sufficient for stability.  In
particular, if $\hat Q$ is normal then it can be unitarily (and
therefore uniformly) diagonalized, $|T(\xi)|=|T^{-1}(\xi)| = 1$.
Since for the time integrators that we consider $\hat Q$ is a
polynomial in $\hat P$, $\hat Q$ will be normal if $\hat P$ is normal
(as would be the case if $\hat{P}$ were Hermitian or anti-Hermitian).
Note that if the von Neumann condition is violated then the scheme is
not stable in any sense.

It is possible for a discretization to be (conditionally) stable
without $\hat Q$ being normal (and hence unitarily diagonalizable).
This turns out to be the case for most systems considered in this
work.  In such cases we find it convenient to introduce the norm
$|\hat u|_{\hat H} = \langle \hat u, \hat H \hat u \rangle^{1/2}$ and
proceed as follows.  Let us assume that $\hat H (\xi)$ are Hermitian
matrices such that
\begin{eqnarray}
&&K^{-1} I \le \hat H (\xi) \le K I,\label{eqn:discrsym}\\
&&| \hat Q |_{\hat H} \le e^{\alpha k},\nonumber
\end{eqnarray}
where $K$ is a positive constant. Notice that\footnote{For a positive
definite Hermitian matrix $H$, $H^{\alpha}$ (for $\alpha$ not
necessarily an integer) is defined as $S^* D^{\alpha} S$ where $H =
S^* D S$ and $D$ is the diagonal matrix of positive real eigenvalues.}
$|\hat u|_{\hat H} = |\hat H^{1/2} \hat u | \le K^{1/2} |\hat u|$ and
$K^{-1} |A| \le |A|_{\hat H} = |\hat H^{1/2} A \hat H^{-1/2} | \le K
|A|$.  As a consequence the von Neumann condition is satisfied,
$\sigma (\hat Q) = \sigma (\hat H^{1/2} \hat Q \hat H^{-1/2}) \le |
\hat H^{1/2} \hat Q \hat H^{-1/2} | = | \hat Q |_{\hat H} \le e^{\alpha k}$,
where $\sigma(\hat Q)$ denotes the spectral radius of $\hat Q$.
Stability follows from
\begin{equation}
| \hat Q^n| \le K | \hat Q^n |_{\hat H} \le K | \hat Q |^n_{\hat H} \le K
  e^{\alpha t_n}. \label{Eq:Qhatn}
\end{equation}

According to the Kreiss Matrix Theorem (Sec.~4.9 of \cite{RM}), for a
family ${\mathcal F}$ of $m\times m$ matrices $A$ the following two
statements are equivalent:
\begin{enumerate}
\item There exists a constant $C$ such that for all $A \in \mathcal F$
  and all positive integers $n$
\[
| A^n | \le C
\]
\item There is a constant $K>0$ and, for each $A\in \mathcal F$, a
  positive definite Hermitian matrix $H$ with the properties
\[
K^{-1} I \le H \le K I, \qquad A^*HA \le H
\]
\end{enumerate}
This implies that condition (\ref{Eq:Qhatn}) is also necessary for
stability.

\subsubsection{Number of grid points}
\label{Sec:oddcase}

In this review we have assumed that the number of grid points in each
direction is even.  This means that no matter how small the number of
grid points is, as long as it is even, the highest frequency $\xi_r =
\pi$ is present.  To allow for an odd number of grid points one must
change the summation range in Eq.~(\ref{Eq:intpol}) to $\omega_r =
-(N-1)/2, \ldots (N-1)/2$, in which case, $|\xi_r|$ never equals $\pi$,
although it does approach this value as $h\to 0$.

\subsection{A sufficient condition for stability}
\label{Sec:sufficient}

We can now give a simpler sufficient condition for numerical
stability.  This condition applies to systems which admit a conserved
energy in Fourier space and will enable us in
Sec.~\ref{Sec:Stabdiscr12} to obtain another condition suitable for
the applications.  We consider only time integrators such that
\begin{equation}
\hat Q = {\mathcal P} (k \hat P)
\end{equation}
The eigenvalues $q_{\nu}$ of $\hat Q$ are related to the eigenvalues
$p_{\nu}$ of $\hat P$ by $q_{\nu} = {\mathcal P} (k p_{\nu})$. This
can be seen by using Shur's lemma.  Provided that the eigenvalues
$p_{\nu}$ are imaginary, the inequality $|q_{\nu}| \le 1$ is
equivalent to $k p_{\nu} \le \alpha_0$, where $\alpha_0 = 2$ for ICN,
$\sqrt{8}$ for 4RK, $\sqrt{3}$ for 3RK.  Hence,
\begin{equation}
\sigma(k \hat P) \le \alpha_0 \label{Eq:sigmakP}
\end{equation}
is equivalent to $\sigma(\hat Q) \le 1$.  Condition (\ref{Eq:sigmakP})
is called {\em local stability on the imaginary axis} in
\cite{KreissScherer}.  Suppose that the time step is such that $\sigma
(k\hat P) \le \alpha_0$.  If we can find Hermitian matrices $\hat
H(\xi)$ such that
\begin{eqnarray}
&&K^{-1} I \le \hat H (\xi) \le K I\,, \label{Eq:discrsymm1}\\
&&\hat H(\xi) \hat P(\xi) + \hat P(\xi)^*\hat H(\xi) = 0 \,,
\label{Eq:discrsymm2}
\end{eqnarray}
we say that $\hat H(\xi)$ is a {\em discrete symmetrizer} of $\hat
P(\xi)$.  The matrices $\hat H^{1/2} \hat P \hat H^{-1/2}$ are
anti-Hermitian, hence they can be diagonalized by unitary matrices
$S(\xi)$.  This implies that the matrices $\hat H^{-1/2}(\xi)
S(\xi)$ diagonalize $\hat Q(\xi)$.  The inequality
\begin{eqnarray}
|\hat Q|_H &=& |\hat H^{1/2} \hat Q \hat H^{-1/2} | = | S^{-1} \hat
 H^{1/2} \hat Q \hat H^{-1/2} S | = \sigma(\hat Q) \le 1
\end{eqnarray}
guarantees stability.  In fact, the amplification matrix can be uniformly
diagonalized by $T(\xi) = \hat H^{-1/2}(\xi) S(\xi)$.

In applications one would construct a norm (i.e., matrices $\hat
H(\xi)$ satisfying (\ref{Eq:discrsymm1})) which is conserved by the
Fourier transformed semi-discrete evolution equations,
\begin{equation}
\frac{d}{dt} |\hat v |^2_{\hat H} = \langle \hat v , (\hat H \hat P + \hat
P^* \hat H) \hat v \rangle  = 0\,.\label{Eq:conservation}
\end{equation}
This implies that condition (\ref{Eq:discrsymm2}) holds and $\hat
H(\xi)$ is a discrete symmetrizer.

To construct $\hat H$ one can proceed as follows.  Assume the
existence of a matrix $T$ such that $T^{-1} \hat P T = \Lambda$ is
diagonal with imaginary elements.  Then the quantity $\hat{v}^* \hat H
\hat{v}$, where $\hat H = T^{-1 *} D T^{-1}$ and $D$ is a positive
definite matrix which commutes with $\Lambda$, is conserved by the
system $\p_t \hat{v} = \hat P \hat{v}$.  Defining the {\em
characteristic variables} of $\hat P$ to be $\hat w \equiv T^{-1} \hat
v$ (these are individually conserved: $\partial_t | \hat w_i|^2 = 0$),
we see that to construct a conserved quantity one can take $\hat w^* D
\hat w$.  (For $D=I$ this corresponds to adding the squared absolute
values of the characteristic variables.)  For $\hat H$ to be a
symmetrizer it remains to be established that $K^{-1} | \hat v |^2 \le
\hat v^* \hat H \hat v \le K | \hat v |^2$.

\section{Stability of first order in time, second order in space systems}
\label{Sec:Stability12}
\label{Sec:DiscrSymm}

What we have done so far applies to fully first order systems.  We
have shown that if inequalities (\ref{Eq:sigmakP}) and
(\ref{Eq:discrsymm1}) and Eq.~(\ref{Eq:conservation}) hold, then the
fully discrete scheme is stable and satisfies the estimate
(\ref{Eq:stability}) with $\alpha = 0$.  In this section we show how
this can be extended to second order in space systems.  We first look
at the continuum problem and then investigate its standard
discretization.

\subsection{Well-posedness of first order in time and second order in
  space hyperbolic systems} 
\label{Sec:Reduction}

It is possible for the Cauchy problem of a first order in time and
second order in space system of equations to be ill-posed in the $L_2$
norm, but well-posed in a norm which contains additional derivatives
(see the introduction).  The system is still useful; for example, a
suitable finite difference approximation of the equations can be
convergent in the discrete $L_2$ norm.  We analyze the well-posedness
of the Cauchy problem for such systems by using the analytical tool of
a {\em reduction to first order}.  This will be done in Fourier space,
so that the number of additional variables being introduced is
minimized \cite{KO}.

Consider system (\ref{Eq:Cauchy1}) with $p=2$ and suppose that it can
be written in the form
\begin{eqnarray}
&&\p_t \bm{u} = P \bm u\,,\qquad 
\bm{u} = \columntwo u v,\label{eqn:redsecord}\\
&& P = 
\matrixtwotwo{A^i \p_i + B}                 {C}
             {D^{ij} \p_i \p_j + E^i\p_i+F} {G^i\p_i+J}
,\nonumber
\end{eqnarray}
where the evolved variables have been split into two types.  The
column vector $u$ represents those that are differentiated twice (in
space) and $v$ represents those that are not.  In $P$ a sum over
repeated indices is assumed.  Not all second order in space systems
can be written in this form (for example, $u_t = u_{xx}$).  This form
is general enough to include all the first order in time, second order
in space systems that we have considered that can be reduced to first
order in space.  Fourier transforming this system, we obtain
\begin{eqnarray}
&&\p_t \hat{\bm{u}} = \hat P \hat{\bm{u}}\,, \qquad \hat{\bm{u}} =
\columntwo \uhat \vhat , \label{eqn:secordsym}\\
&&\hat P = 
\matrixtwotwo{i \omega A^n + B}                 {C}
             {- \omega^2 D^{nn} + i \omega E^n+F} {i \omega G^n + J}
,\nonumber
\end{eqnarray}
where $M^n \equiv M^i n_i$ and $\omega_i \equiv |\omega| n_i$ and
$\omega \equiv |\omega|$.  We define the {\em second order principal
symbol} to be
\begin{eqnarray}
  \hat P' = 
  \matrixtwotwo    {i \omega A^n}      {C}
                   {- \omega^2 D^{nn}} {i \omega G^n}
.\label{eqn:pprime}
\end{eqnarray}

We now state the main result of this subsection.  If there exists
$\hat H(\omega) = \hat H^*(\omega)$ such that the {\em energy}
$\hat{\bm{u}}^* \hat H \hat{\bm{u}}$ is conserved by the principal
system $\p_t \hat{\bm{u}} = \hat P' \hat{\bm{u}}$ and $\hat H$
satisfies
\begin{eqnarray}
  K^{-1} I_\omega \le  \hat H  \le K I_\omega,
 \quad I_\omega \equiv 
  \matrixtwotwo    {\omega^2}      {0}
                   {0} {1}
,\label{eqn:redScondition}
\end{eqnarray}
where $K$ is a positive scalar constant, then the solution of
(\ref{eqn:redsecord}) satisfies the estimate
\begin{eqnarray}
&&\|\bm{u}(t,\cdot)\| \le K e^{\alpha t} \|\bm{u}(0,\cdot)\|,\label{eqn:redestimate}\\
&&\|\bm{u}\|^2 \equiv \int |u|^2 + \sum_{i=1}^d|\p_i u|^2 + |v|^2 d^dx\,, \nonumber
\end{eqnarray}
and the problem is well-posed in this norm\footnote{Note that we made
  no assumptions regarding the smoothness of the matrix $\hat
  H(\omega)$.  In view of generalizations of this work to the variable
  coefficient case it may be desirable to demand that $T^{-1*}\hat H
  T^{-1}$, where $T$ is defined in Eq.~(\ref{Eq:T}), be smooth in all
  variables.}.

The proof proceeds via a pseudo-differential reduction to first order
\cite{NOR}.  This involves the introduction of a new variable $\what =
i \omega \uhat$.  By taking a time derivative of this definition, we
obtain the enlarged system in which the second derivative of $\uhat$
has been replaced with a first derivative of $\what$. We reduce the
order of the system as much as possible so that any occurrence of $i
\omega \hat u$ is replaced with $\hat w$. 
This particular first order reduction is
\begin{eqnarray}
&&\p_t \hat{\bm{u}}_R = \hat P_R \hat{\bm{u}}_R\,,\qquad \hat{\bm{u}}_R
= \columnthree \uhat \what \vhat ,\label{eqn:forwithlot}\\
&&\hat P_R = 
\matrixthreethree{B} {A^n}                   {C}
                 {0} {i \omega A^n + B}      {i \omega C}
                 {F} {i \omega D^{nn} + E^n} {i \omega G^n + J}
.\nonumber
\end{eqnarray}

This system is equivalent to the second order system
(\ref{eqn:secordsym}) only when the {\em auxiliary constraints}
\begin{equation}
\hat C(t,\omega) \equiv \hat w(t,\omega) - i \omega \hat u(t,\omega) = 0 \label{Eq:reductconstraint}
\end{equation}
are satisfied.  It can be shown that $\partial_t \hat C = B \hat C$ so if these
constraints are satisfied initially, then they are satisfied for all
time.  They are said to be {\em propagated} by the first order
evolution equations.

If this system admits a matrix $\hat H_R$ satisfying
(\ref{eqn:contsym}) then the solutions satisfy the estimates
\begin{eqnarray}
|\hat{\bm{u}}_R(t,\omega)| \le K e^{\alpha t} |\hat{\bm{u}}_R(0,\omega)| \,,
\label{Eq:KLestimate}
\end{eqnarray}
where $|\hat{\bm{u}}_R|^2 \equiv |\uhat|^2 + |\what|^2 +
|\vhat|^2$, for arbitrary initial data and $\omega$.  Specifically,
the estimate holds for solutions which satisfy the auxiliary
constraints and therefore correspond to solutions of the second order
system.  The uniform estimate in $\omega$ of
\begin{eqnarray}
|\uhat|^2 + \omega^2 |\uhat|^2 + |\vhat|^2 = 
|\uhat|^2 +  \sum_{i=1}^d |i \omega_i  \uhat|^2 + |\vhat|^2
\end{eqnarray}
implies, by Parseval's relation, the estimate in real space
\begin{eqnarray}
&&\| \bm{u}(t,\cdot)\| \le K e^{\alpha t} \| \bm{u}(0,\cdot)\|\,,\\
&&\|\bm{u}\|^2 \equiv \int |u|^2 + \sum_{i=1}^d | \partial_i u|^2 + |v|^2  d^dx\,.\nonumber
\end{eqnarray}
So the existence of $\hat H_R$ for a first order
pseudo-differential reduction implies the well-posedness of the
second order system with respect to a norm containing derivatives.

We have still to show that we can find an $\hat H_R$ for
(\ref{eqn:forwithlot}).  Whether or not this is the case is
independent of the {\em lower order} terms $\hat P_R$ contains.  A
calculation similar to Lemma 2.3.5 in \cite{KL-Book} shows that if
$\hat P(\omega)$ admits an $\hat H_R$, then so will $\hat P(\omega) +
B(\omega)$, where $B(\omega)$ is any matrix which satisfies $|B|
+|B^*| \le C$ for $C$ independent of $\omega$.  In other words, the
terms that are not multiplied by $i\omega$ can be dropped from
(\ref{eqn:forwithlot}), giving the principal symbol of the first order
reduction
\begin{eqnarray}
\hat P_R' = 
\matrixthreethree{0} {0}                   {0}
                 {0} {i \omega A^n}      {i \omega C}
                 {0} {i \omega D^{nn}} {i \omega G^n}
\label{eqn:PRprimed}
\end{eqnarray}
without affecting the well-posedness.  The principal symbols of the
second order system, Eq.~(\ref{eqn:pprime}), and the first order
pseudo-differential reduction, Eq.~(\ref{eqn:PRprimed}), are related by
\begin{eqnarray}
\hat P_R' = \matrixtwotwo {0}  {0}
                          {0}  {T \hat P' T^{-1}}
,\qquad
 T \equiv 
\matrixtwotwo    {i\omega}      {0}
                 {0} {1}
. \label{Eq:T}
\end{eqnarray}
(Note that $T^{-1}$ does not exist for $\omega = 0$.  However, in this
case, $\hat P_R'=0$, and admits the identity as a symmetrizer.) By
assumption, there exists $\hat H(\omega) = \hat H^*(\omega)$ such that
$\hat{\bm{u}}^* \hat H \hat{\bm{u}}$ is conserved by the principal
system $\p_t \hat{\bm{u}} = \hat P' \hat{\bm{u}}$ and satisfies
(\ref{eqn:redScondition}).  This $\hat H$ satisfies $\hat H \hat P' +
\hat P'^* \hat H = 0$, and it is straightforward to show that
\begin{equation}
\hat H_R \equiv \matrixtwotwo 1 0
                         0 {T^{-1*} \hat H T^{-1}}
\end{equation}
satisfies $\hat H_R = \hat H^*_R$ and $\hat H_R \hat P_R' + \hat
P_R'^* \hat H_R = 0$.  Further, by noting that $T^* T = I_\omega$,
using (\ref{eqn:redScondition}) one can show that $\hat H_R$ satisfies
$K^{-1}I \le \hat H_R \le KI$.  Hence we have found a symmetrizer of
$\hat P_R'$ and the result has been proved\footnote{It can also be
shown that $\hat P'_R$ is diagonalizable with the same eigenvalues as
$\hat P'$, plus as many zeroes as there are components of $u$.}.

To construct $\hat H$ one can use the characteristic variables of
$\hat P'$, as described at the end of Sec.~\ref{Sec:sufficient}.  We
would like to point out that this analysis did not require that the
auxiliary constraint propagation problem be well-posed.  These
constraints are merely a tool for the analysis of the system.  We only
need to establish uniqueness of the solution with zero initial data
for the auxiliary constraints.  In the linear constant coefficient
case this result is trivial.  When evolving the second order system,
these constraints are identically zero at all times.  An alternative
to the pseudo-differential reduction method is to perform a fully
differential reduction by introducing a new variable in physical space
for each derivative (see, for example, \cite{SCPT,BS}).

\subsection{Stability of discretizations of first order in time and
  second order in space systems}
\label{Sec:Stabdiscr12}

We now show how the continuum analysis of the previous subsection can
be extended to the fully discrete case.  The semi-discrete finite
difference approximation of (\ref{eqn:redsecord}) can be written as
\begin{eqnarray}
&&\frac d {dt} \bm{v} = P \bm{v},\qquad \bm{v} = \columntwo u v , \label{eqn:redsecorddisc}\\
&& P = 
\matrixtwotwo{A^i \D_i + B}                 {C}
             {D^{ij} \DD_{ij} + E^i \D_i + F} {G^i \D_i+J}
,\nonumber
\end{eqnarray}
where $\D_{i}$ is a discretization of the first derivative in the $i$
direction and $\DD_{ij}$ is a discretization of the second derivative in
the $i$ and $j$ directions.  For example, the standard second order
accurate discretization would have
\begin{eqnarray}
\D_i = D_{0i}, \quad
D^{(2)}_{ij} = \left\{ \begin{array}{l@{\qquad}l}
D_{0i}D_{0j} & i\neq j\\
D_{+i}D_{-i} & i = j
\end{array}\right. .
\end{eqnarray}
The principal symbol of the semi-discrete system is 
\begin{eqnarray}
  \hat P' = 
  \matrixtwotwo    {A^i \Dhat_i}      {C}
                   {D^{ij} \DDhat_{ij} } {G^i \Dhat_i}
,
\end{eqnarray}
where 
\begin{equation}
\Dhat_i = \frac{i}{h} \sin{\xi_i},\quad 
\DDhat_{ij} = \left\{ \begin{array}{l@{\qquad}l}
-\displaystyle\frac{1}{h^2}\sin{\xi_i}\sin{\xi_j} & i\neq j\\
-\displaystyle\frac{4}{h^2}\sin^2{\frac{\xi_i}{2}} & i = j
\end{array}\right. ,
\end{equation}
for the standard second order discretization.  The {\em pseudo-discrete}
first order reduction is obtained by defining
\begin{eqnarray}
\what \equiv i \Omega \uhat\,,\qquad  \Omega^2 = \sum_{i=1}^d |\hat D_{+i}|^2.
\end{eqnarray}
The reduced system is
\begin{eqnarray}
&&\frac d {dt} \hat{\bm{v}}_R = \hat P_R \hat{\bm{v}}_R,\qquad
  \hat{\bm{v}}_R = \columnthree \uhat \what \vhat, \label{eqn:forwithlotdisc}\\
&& \hat P_R = 
\matrixthreethree{B} {(i \Omega)^{-1} A^i \Dhat_i}                   {C}
                 {0} { A^i \Dhat_i  + B}      {i \Omega C}
                 {F} {(i \Omega)^{-1} (D^{ij} \DDhat_{ij} + E^i \Dhat_i)}
  {G^i \Dhat_i + J} .
\label{eqn:PR}
\end{eqnarray}
We can show that the discrete auxiliary constraint is preserved by the
time integrator.  Define $c \equiv ( {-i\Omega I} \; I \; 0 )$, so
that the constraint is $c \hat{\bm{v}}_R = \what - i \Omega \hat u=
0$.  Since $c \hat P_R \hat{\bm{v}}_R = B c \hat{\bm{v}}_R$, we have
that $c \hat{\bm{v}}_R = 0$ implies $c \hat P_R \hat{\bm{v}}_R = 0$
and hence $c \hat P_R^n \hat{\bm{v}}_R = 0$ and $c {\mathcal P}(k \hat
P) \hat{\bm{v}}_R = 0$.  Now consider evolving the reduced system with
a polynomial time integrator; i.e. $\hat{\bm{v}}_R^{n+1} = {\mathcal
P}(k \hat P_R) \hat{\bm{v}}_R^n$.  If the auxiliary constraints are
satisfied on one time step, then they are satisfied on the next as
well, since $c \hat{\bm{v}}^n_R = 0$ implies $c \hat{\bm{v}}_R^{n+1} =
c {\mathcal P}(k \hat P) \hat{\bm{v}}_R^n = 0$.  Hence there is a
one-to-one correspondence between solutions of the second order fully
discrete system and those of the constraint-satisfying reduced
system. Note that we have used the fact that the time integrator is a
polynomial in $\hat P_R$, as is the case for systems with constant
coefficients.  This result can be extended to the variable coefficient
case, where one would have to perform the reduction to first order in
physical space by introducing the gridfunctions $X^{(i)} = D_{+i}u$.

Making use of Theorem 5.1.2 of \cite{GKO-Book}, the terms which
correspond to the continuum lower order terms can be dropped from
$\hat P_R$ without affecting the stability of the fully discrete
system, provided that $(i\Omega)^{-1} \Dhat_i$, $k\Dhat_i$ and
$k\Omega^{-1} \DDhat_{ij}$ are bounded.  This guarantees that the
assumptions of the theorem are satisfied.  This is true for the second
and fourth order accurate standard discretizations.

The result for stability of the fully discrete problem is analogous to
that for well-posedness at the continuum.  If there exists $\hat H(\xi) =
\hat H^*(\xi)$ such that the energy $\hat{\bm{v}}^* \hat H \hat{\bm{v}}$ is
conserved by the semi-discrete principal system $\p_t \hat{\bm{v}} =
\hat P' \hat{\bm{v}}$ and $\hat H$ satisfies
\begin{eqnarray}
  K^{-1} I_\Omega \le  \hat H  \le K I_\Omega, \quad I_\Omega \equiv 
  \matrixtwotwo    {\Omega^2}      {0}
                   {0} {1}
,  \label{Eq:dequiv}
\end{eqnarray}
where $K$ is a positive scalar constant, then it is possible to
construct a discrete symmetrizer for the first order reduction with no
lower order terms.  So if, in addition, the principal symbol $\hat P'$
satisfies $\sigma(k \hat P') \le \alpha_0$, the fully discrete system
(including lower order terms) is stable with respect to the norm
\begin{eqnarray}
\|\bm{v}\|_{h,D_+}^2 \equiv \|u\|_h^2 + \|v\|_h^2 +
\sum_{i=1}^d\|D_{+i} u\|_h^2, \label{Eq:stabnorm} 
\end{eqnarray}
i.e. the solution satisfies the estimate
\begin{eqnarray}
\|\bm{v}^n\|_{h,D_+} \le K e^{\alpha t_n}
\|\bm{v}^0\|_{h,D_+}\,. \label{Eq:Dpestimate} 
\end{eqnarray}

Again, $\hat H$ can be constructed from the characteristic variables
of $\hat P'$, as described at the end of Sec.~\ref{Sec:sufficient}.
Note that the matrix $\hat P_R$ is not defined for $\Omega=0$.
However, this does not cause any difficulties in the linear constant
coefficient case.  One can write the space of solutions as a direct
sum consisting of constant functions plus a space of solutions with
nontrivial $\Omega$, and treat each subspace independently.

\subsection{Convergence}
We briefly discuss convergence of the solution of the discrete problem
to that of the continuum problem.  We assume that
(\ref{Eq:Dpestimate}) holds.  Inserting the exact smooth solution
$\bm{u}(t,x)$ into the scheme $\bm{v}^{n+1} = Q \bm{v}^n$ generates
truncation errors as inhomogeneous terms in the difference
approximation and in the initial data.  The error grid-function
$\bm{w}^n_j \equiv \bm{v}^n_j - \bm{u}(t_n,x_j)$ satisfies
\begin{eqnarray}
\bm{w}^{n+1}_j &=& Q \bm{w}^n_j + \bm{\tilde F}^n_j\,, \label{Eq:insexact1}\\
\bm{w}^0_j &=& \bm{\tilde f}_j\,, \label{Eq:insexact2}
\end{eqnarray}
where $\bm{\tilde F}^n_j = \bm{\phi}(t_n,x_j) O(k^{p_1} + h^{p_2})$, and
$\bm{\tilde f}_j = \bm{\psi}(x_j) O(h^{p_3})$ with $\bm{\phi}$ smooth.  The temporal accuracy
of the scheme is $p_1$ and the spatial accuracy is $p_2$.  The
discrete version of Duhamel's principle (see Theorem 5.1.1 in
\cite{GKO-Book}) gives the estimate
\begin{eqnarray}
\|\bm{w}^n\|_{h,D_+} &\le& K e^{\alpha t_n} \left( \| \bm{w}^0 \|_{h,D_+} + k
 \sum_{r=0}^{n-1} \| \bm{\tilde F}^r\|_{h,D_+}\right) \le O(k^{p_1} +
 h^{p_2})\,, 
\label{Eq:errorestimate}
\end{eqnarray}
provided that the initial data satisfies $\| \bm{w}^0 \|_{h,D_+} \le
O(h^{p_2})$.  If $\bm{\psi}$ is smooth this condition is
satisfied and, in particular, it is satisfied for exact initial data.

Inequality (\ref{Eq:errorestimate}) implies convergence with respect
to the discrete $L_2$ norm, $\|\bm{w}\|_h \le \| \bm{w}\|_{h,D_+}$,
despite the scheme being unstable with respect to this norm.  Note
that $p$-th order convergence is obtained, with $p = \min (p_1,p_2)$
assuming $k=\lambda h$, even though the norm contains first order
accurate one-sided difference operators.

\section{Applications}
\label{Sec:Applications}

In the following subsections we apply the theoretical tools discussed
in Sec.~\ref{Sec:Stability12} to different systems.  We start with a
first order strongly hyperbolic system with no lower order terms.  We
then investigate three second order in space systems: the wave
equation, a generalization of the KWB formulation of Maxwell's
equations and the NOR formulation of Einstein's equations.  We show
that the clear correspondence between strong hyperbolicity and the
existence of a discrete symmetrizer which occurs in first order
systems with no lower order terms is lost when the standard
discretization is used for second order in space systems.  Similarly,
the simple correspondence between characteristic speeds and the von
Neumann condition, Eq. (\ref{Eq:stablimFOSH}), does not hold for
second order in space systems.  It is convenient to define the
following quantities,
\begin{equation}
\chi^2_q = \sum_{i=1}^d \sin^q\frac{\xi_i}{2}\,,\qquad \chi^2 = \sum_{i=1}^d
\sin^2\xi_i\,,\qquad \Omega = \frac{2\chi_2}{h}\,,
\end{equation}
Note that the maximum of $\chi_q$ and $\chi$ is $\sqrt{d}$.  We also
recall that when the eigenvalues of $\hat P$ are imaginary,
\begin{equation}
\sigma(k \hat P) \le \alpha_0 \iff \sigma(\hat Q) \le 1\,,
\end{equation}
where $\alpha_0 = 2$ for ICN, $\sqrt{8}$ for 4RK and $\sqrt{3}$ for
3RK.

\subsection{Stability of first order strongly hyperbolic systems}

Our first application is a constant coefficient first order system in
$d$ spatial dimensions
\begin{equation}
\frac{\p u}{\p t} = \sum_{i=1}^d A^{i} \frac{\p u}{\p x^i}\,, \label{Eq:FOSH}
\end{equation}
where $u$ is a vector valued function of the space-time coordinates.
We assume that the system is strongly hyperbolic and that it admits a
symmetrizer, i.e., there
exists a matrix $\hat H(\omega)$ in Fourier space, such that $\hat
H(\omega) \hat P (i \omega) + \hat P^{*}(i\omega) \hat H(\omega) = 0$,
where $\hat P (i\omega) = i \sum_{i=1}^d\omega_i A^i$.
The discrete symbol associated with the standard second order accurate
discretization of this system is
\[
\hat P_h(\xi) = \frac{i}{h} \sum_{i=1}^d A^{i}\sin\xi_i = \hat P(i h^{-1} \sin \xi) \,,
\]
where we attached the subscript $h$ to the discrete symbol to
distinguish it from that of the continuum.  We now construct the
discrete symmetrizer
\begin{equation}
\hat H_h(\xi) \equiv \hat H(h^{-1} \sin \xi) \,.
\label{Eq:twoH}
\end{equation}

Conditions (\ref{Eq:discrsymm1})--(\ref{Eq:discrsymm2}) are satisfied
and condition (\ref{Eq:sigmakP}) is sufficient for stability.  The
latter becomes $\sigma(k\hat P) = \lambda \chi \sigma(A(n)) \le
\alpha_0$, where $A(n) = \sum_{i=1}^d n_i A^{i}$, $n_i = \chi^{-1}
\sin\xi_i $, so that $\sum_{i=1}^d n_i^2 = 1$.  Since this inequality
must hold for all $\xi_i$, and the quantity $\chi$ reaches its maximum
value $\sqrt{d}$ at $\xi_i = \pm \pi / 2$, we obtain the stability
condition
\begin{equation}
\lambda \le \frac{\alpha_0}{\sigma(A(n)) \sqrt{d}}\,. \label{Eq:stablimFOSH}
\end{equation}
In the symmetrizable hyperbolic case one can take the discrete
symmetrizer to be the same as that of the continuum (which, by definition, is
independent of $\omega$)
\begin{equation}
\hat H_h(\xi) = H. \label{Eq:symmfirst}
\end{equation}  

This analysis of first order strongly hyperbolic systems shows
that if the characteristic speeds depend neither on the direction nor
on the dimensionality of the problem, i.e., if $\sigma(A(n))$ depends
neither on $n$ nor on $d$, then the Courant limit has a $1/\sqrt{d}$
dependence.  In addition, when the second order accurate centered
difference operator $D_0$ is used to approximate the spatial
derivatives, a Courant limit violation would manifest itself as a
rapid growth of the mid high frequency mode $|\xi_i| = \frac{\pi}{2}
\approx 1.571$.  This mode is present if $N$ is a multiple of $4$.  A
similar analysis shows that in the fourth order accurate case the
situation differs.  The Courant limit is $1.372$ times smaller than
(\ref{Eq:stablimFOSH}) and above this limit the most rapid growth
occurs at a slightly higher frequency, $|\xi_i| = 2 \arctan (6^{1/2} /
(4 - 6^{1/2}))^{1/2} \approx 1.797$.  See also Appendix \ref{Sec:NumProp12}.

\subsection{First order in time and second order in space wave equation}
\label{Sec:Wave}

In this section we discuss the stability properties of an
approximation of the $d$ dimensional wave equation written as a first
order in time and second order in space system
\begin{eqnarray}
\p_t \phi(t,x) &=& \Pi(t,x)\,, \label{Eq:wave1}\\
\p_t \Pi(t,x) &=& \sum_{i=1}^d \p_i^2\phi(t,x)\,. \label{Eq:wave2}
\end{eqnarray}
In the introduction we pointed
out that the Cauchy problem for this system is not well-posed in
$L_2$.  One can expect that a direct application of definition
(\ref{Eq:stability}), which is based on the discrete $L_2$ norm, to a
scheme approximating (\ref{Eq:wave1})--(\ref{Eq:wave2}) would lead to
the conclusion that the scheme is unstable. The first order reduction,
however, is well-posed in $L_2$ (it is symmetric hyperbolic), hence
the second order system satisfies an energy estimate with respect to
\begin{equation}
\| \bm{u}(\cdot, t) \|^2 = \int |\phi(x,t)|^2 + |\Pi(x,t)|^2 +
\sum_{i=1}^d |\p_i \phi(x,t)|^2  \ud^d x\,.
\end{equation}
In this section we show stability for the standard discretization of
this system, both by the pseudo-discrete reduction method given in
Sec.~\ref{Sec:Stabdiscr12}, and by a direct discrete reduction
in physical space. The two methods give equivalent results.

Following the method of lines, we first discretize space and leave
time continuous,
\begin{eqnarray}
\frac{d}{dt} \phi_j(t) &=& \Pi_j(t)\,, \label{Eq:semiwave1}\\
\frac{d}{dt} \Pi_j(t) &=& \sum_{i=1}^d D_{+i}D_{-i} \phi_j(t)\,.\label{Eq:semiwave2}
\end{eqnarray}
Using the technique described in Sec.~\ref{Sec:Stabdiscr12}, we
see that the (principal) symbol of the second order semi-discrete
problem
\begin{equation}
\hat P = \left( \begin{array}{cc} 0 & 1\\ -\Omega^2 & 0
\end{array} \right),\qquad
T^{-1} = \left( \begin{array}{cc} i\Omega & 1\\ -i\Omega & 1
\end{array} \right),
\end{equation}
has purely imaginary eigenvalues $\pm i\Omega$.  The matrix $T$
diagonalizes $\hat P$. 
The sum of the squared moduli of the characteristic variables gives
the conserved energy (here $D=1/2 I$)
\begin{eqnarray}
\hat{\bm{v}}^* (T^{-1})^* D T^{-1} \hat{\bm{v}} \equiv |i\Omega \hat \phi|^2 +
|\hat \Pi|^2 = \Omega^2 |\hat \phi|^2 + |\hat \Pi|^2.
\end{eqnarray}
By taking $K=1$ in (\ref{Eq:dequiv}) we see that we have numerical
stability with respect to the discrete norm
\begin{equation}
\| \bm{v} \|^2_{h,D_+} = \sum_j (\phi_j^2+\Pi_j^2 + \sum_{i=1}^d (D_{+i}
  \phi_j)^2) h^d, \label{Eq:denergywave}
\end{equation}
provided that the von Neumann condition
\begin{equation}
\lambda \le \alpha_0 / (2 \sqrt{d})\,, \label{Eq:vNwave}
\end{equation}
which follows from $\sigma (k\hat P) = k\Omega = 2\lambda \chi_2 \le
\alpha_0$, is satisfied.

We now illustrate a different method for proving stability of this
system.  A {\em discrete reduction to first order} can be performed
before going to Fourier space.  We introduce the quantities
\begin{equation}
X^{(i)}_j = D_{+i} \phi_j\label{Eq:X}
\end{equation}
and obtain the reduced system
\begin{eqnarray}
\frac{d}{dt} \phi_j(t) &=& \Pi_j(t)\,,\label{Eq:red1}\\
\frac{d}{dt} \Pi_j(t) &=& \sum_{i=1}^d D_{-i} X^{(i)}_j(t)\,,\label{Eq:red2}\\
\frac{d}{dt} X^{(i)}_j(t) &=& D_{+i} \Pi_j(t)\,.\label{Eq:red3}
\end{eqnarray}

Notice that only if Eq.~(\ref{Eq:X}) is identically satisfied is the
reduced system equivalent to the original one.  It is important to
check whether the evolution equations (\ref{Eq:red1})--(\ref{Eq:red3})
are compatible with this requirement.  Let $C^{(i)}_j(t) \equiv
X^{(i)}_j - D_{+i}\phi_j$.  If we prescribe initial data such that
$C^{(i)}_j(0) = 0$, then at later times $C^{(i)}_j(t) = 0$.  This is a
consequence of the fact that
\begin{equation}
\frac{d}{dt} C^{(i)}_j (t) = \frac{d}{dt} (X^{(i)}_j(t) -  D_{+i}\phi_j(t)) = 0\,.
\end{equation}
There is a one-to-one correspondence between solutions of
(\ref{Eq:semiwave1})--(\ref{Eq:semiwave2}) and those of
(\ref{Eq:X})--(\ref{Eq:red3}).  Furthermore, one can check that the
time integrator does not spoil the propagation of the constraints.

Ignoring lower order terms, the symbol associated with the reduced
system (\ref{Eq:red1})--(\ref{Eq:red3}) is anti-Hermitian, therefore
Eq.~(\ref{Eq:discrsymm2}) is satisfied with $\hat H=1$.  The
non-trivial eigenvalues of $\hat P$ are $\pm i \Omega$, the
same as those of the original system
(\ref{Eq:semiwave1})--(\ref{Eq:semiwave2}).  This proves that
(\ref{Eq:vNwave}) is a necessary and sufficient condition for
stability with respect to the discrete norm (\ref{Eq:denergywave}).

This specific discrete reduction to first order, and the
pseudo-discrete reduction to first order described in
Sec.~\ref{Sec:Stabdiscr12} give equivalent results.

\subsubsection{Fourth order accuracy}

In hyperbolic problems a fourth order accurate spatial discretization
requires significantly fewer grid-points per wavelength for a given
tolerance error (see \cite{GKO-Book} and appendix
\ref{Sec:NumProp12}).  The stability proof for the fourth order accurate
discretization of the $d$-dimensional wave equation
\begin{eqnarray}
\frac{d}{dt} \phi_j(t) &=& \Pi_j(t)\,,\\
\frac{d}{dt} \Pi_j(t) &=& \sum_{i=1}^d D_{+i}D_{-i} \left(1-\frac{h^2}{12} D_{+i}D_{-i} \right)\phi_j(t)
\end{eqnarray}
is similar to the second order accurate case.  The discrete symbol and
diagonalizing matrix are
\begin{equation}
\hat P = \left( \begin{array}{cc} 0 & 1\\ -\Delta^2 & 0
\end{array} \right),\qquad T^{-1} = \left( \begin{array}{cc}
  i\Delta & 1 \\ 
  -i\Delta & 1 \end{array} \right),
\end{equation}
where $\Delta^2 = \frac{4}{h^2}\sum_{i=1}^d \sin^2\frac{\xi_i}{2}\left( 1 +
\frac{1}{3} \sin^2\frac{\xi_i}{2}\right)$, has purely imaginary
eigenvalues $\pm i \Delta$.  Taking $D = 1/2 I$ we get the conserved quantity
\begin{equation}
(T^{-1} \hat{\bm{v}})^* D \hat
T^{-1} \hat{\bm{v}}  = \Delta^2 |\hat \phi|^2 + |\hat \Pi|^2.
\end{equation}
Since $\Omega^2 \le \Delta^2 \le \frac{4}{3} \Omega^2$, by taking
$K=4/3$ in (\ref{Eq:dequiv}) we see that we have numerical stability
with respect to the norm (\ref{Eq:denergywave}) provided that the
principal symbol $\hat P$ satisfies $\sigma(k \hat P) \le \alpha_0$.
This gives a stability limit of $\lambda \le \sqrt{3} \alpha_0
/(4\sqrt{d})$.

\subsubsection{A note about the $D_0$-norm and the $D_0^2$ discretization}
\label{Sec:D02}

Replacing the one sided difference operators $D_{+i}$ with centered
difference operators $D_{0i}$ in the norm (\ref{Eq:denergywave}) leads
to difficulties, as the $D_0$-norm does not capture the highest
frequency mode.  In fact, it is possible to construct a family of
solutions of (\ref{Eq:semiwave1})--(\ref{Eq:semiwave2}) proportional
to $(-1)^j$ for which the $D_0$-energy estimate fails.  For this
purpose it is sufficient to consider $\phi_j(t) = (-1)^j \cos(2t/h)$,
$\Pi_j(t) = -2/h(-1)^j\sin(2t/h)$, which gives
\begin{equation}
\frac{\| \bm{v}(t)\|_{h,D_0}}{\| \bm{v}(0) \|_{h,D_0}} = \left(
\cos^2\frac{2t}{h} + \frac{4}{h^2} \sin^2\frac{2t}{h}\right)^{1/2}\,,
\end{equation}
where $\| \bm{v}(t) \|^2_{h,D_0} =\sum_j
(\phi_j^2+\Pi_j^2 + (D_0 \phi_j)^2) h$.  It it not
possible to find constants $K$ and $\alpha$ such that the ratio is
bounded by $Ke^{\alpha t}$, independently of the space step $h$.

It has been suggested that the use of $D_0^2$ rather than $D_+D_-$ for
the second spatial derivatives may improve the stability properties of
a second order in space scheme \cite{Bonetal,BabSziWin}.
To investigate this we study the wave equation in one space dimension
discretized as
\begin{eqnarray}
\frac{d}{dt} \phi_j(t) &=& \Pi_j(t)\,, \label{Eq:waveD01}\\
\frac{d}{dt} \Pi_j(t) &=& D_0^2 \phi_j(t)\,. \label{Eq:waveD02}
\end{eqnarray}
The eigenvalues of $k\hat{P}$ are $\pm i\lambda \sin\xi$, which shows
that the von Neumann condition is satisfied as long as $\lambda \le
\alpha_0$.  Both the stencil and the maximum time step compatible with
the von Neumann condition are twice what they are for the $D_+D_-$
discretization.  However, for a given spatial resolution the numerical
speed of propagation has an error which is four times that of the
$D_+D_-$ case (see Appendix \ref{Sec:NumProp12}).

So far, we have only shown that the scheme is unstable if $\lambda >
\alpha_0$.  By looking at the discrete symbol
\begin{equation}
\hat{P}(\xi) = \left(
\begin{array}{cc}
0 & 1\\
-\frac{1}{h^2}\sin^2\xi & 0
\end{array}
\right) \label{Eq:symbD02}
\end{equation}
we see that there might be a problem for $|\xi| = \pi$.  In this case
the symbol is not diagonalizable.  To explicitly show that the system
(\ref{Eq:waveD01})--(\ref{Eq:waveD02}) is unstable with respect to the
norm
\begin{equation}
\| \bm{v} \|_{h,D_+}^2 = \sum_j \left(\phi_j^2 + \Pi_j^2 + (D_+\phi_j)^2\right) h \label{Eq:Dpenergy}
\end{equation}
it is sufficient to consider the family of initial data $\phi_j(0) =
0, \Pi_j(0) = (-1)^j$, generating the solution $\phi_j(t) = (-1)^jt,
\Pi_j(t) = (-1)^j$.  As $h\to 0$ the ratio 
\begin{equation}
\frac{\| \bm{v}(t) \|_{h,D_+}}{ \| \bm{v}(0)\|_{h,D_+}} = \left(1 + t^2 +
 \frac{4t^2}{h^2}\right)^{1/2}
\end{equation}
grows without bound.

Had we chosen the $D_0$-norm, however, we would have concluded that
the scheme satisfies the required estimate.  This is because this norm
does not capture the highest frequency mode $\phi_j = (-1)^j$.  A
desirable requirement of a norm is that if a scheme is stable with
respect to that norm, then it will remain stable with respect to the
same norm when perturbed with lower order terms (independently of how
these are discretized).  The modified problem
\begin{eqnarray}
\frac{d}{dt} \phi_j(t) &=& \Pi_j(t)\,,\\
\frac{d}{dt} \Pi_j(t) &=& D_0^2 \phi_j(t) - D_+\phi_j(t)
\end{eqnarray}
admits the family of exponentially growing solutions $\phi_j(t) =
(-1)^j \exp ( \sqrt{2/h} t)$, $\Pi_j(t) = (-1)^j \sqrt{2/h}
\exp ( \sqrt{2/h} t)$ which leads to unbounded growth in the
ratio 
\begin{equation}
\frac{\| \bm{v}(t) \|_{h,D_0}}{\| \bm{v}(0)\|_{h,D_0}} = \exp \left(\sqrt{\frac{2}{h}}t\right).
\end{equation}
If we want to be able to decide whether a scheme is stable
or not just by looking at the principal part of the discrete system,
then we must conclude that the $D_0$-energy is not a
suitable energy.

We note that the requirement that stability should not depend on how
lower order terms are discretized was crucial.  If we restrict
ourselves to the perturbation $D_0\phi_j$, then the scheme is still
stable with respect to the $D_0$-energy.  If one wants to be able to
discretize lower order terms freely, as we do, then one is forced to
reject the $D_0^2$ discretization.

Clearly it is the presence of high frequency modes that makes the
$D_0^2$ discretization unstable with respect to the $D_+$-norm.  The
introduction of a mechanism that damps high frequency modes, such as
artificial dissipation, may restore stability.  In the system
\begin{eqnarray*}
\frac{d}{dt} \phi_j &=& \Pi_j - \sigma h^3 (D_+D_-)^2 \phi_j\,,\\
\frac{d}{dt} \Pi_j &=& D_0^2 \phi_j - \sigma h^3 (D_+D_-)^2 \Pi_j
\end{eqnarray*}
the same family of initial data used to prove instability of
(\ref{Eq:waveD01})--(\ref{Eq:waveD02}) gives $\|
\bm{v}(t)\|_{h,D_+}/\|\bm{v}(0)\|_{h,D_+} = (1 + t^2 + 4t^2/h^2)^{1/2}
e^{-16\sigma t/h}$, which does not grow without bound.

\subsection{The generalized Knapp-Walker-Baumgarte system}
\label{Sec:KWB}

We now investigate more complex systems.  We adopt the Einstein
summation convention.  We consider the KWB formulation of Maxwell's
equations \cite{KWB}
\begin{eqnarray}
\p_t A_i &=& - E_i\,, \label{Eq:KWB1}\\
\p_t E_i &=& -\p^k\p_k A_i + \p_i \Gamma \,,\label{Eq:KWB2}\\
\p_t \Gamma &=& 0\,,\label{Eq:KWB3}
\end{eqnarray}
and generalize it by introducing $G= \Gamma - r\p^kA_k$, giving
\begin{eqnarray}
\p_t A_i &=& - E_i \,,\label{Eq:gKWB1}\\
\p_t E_i &=& -\p^k\p_k A_i + r \p_i \p^kA_k + \p_i G \,,\label{Eq:gKWB2}\\
\p_t G &=& r \p^k E_k \,. \label{Eq:gKWB3}
\end{eqnarray}
For $r = 0$ we recover (\ref{Eq:KWB1})--(\ref{Eq:KWB3}) and for $r=1$
we obtain the Z1 system \cite{Z1}, which was recently introduced as a
toy model for the Z4 formulation of General Relativity (see
Sec.~\ref{Sec:Z4}).  We will show that although the parameter $r$
plays no role at the continuum, at the discrete level it can have a
severe impact on the stability properties.

\subsubsection{Continuum analysis}

If we Fourier transform (\ref{Eq:gKWB1})--(\ref{Eq:gKWB3}) and
introduce $\hat \Gamma = \hat G + r i \omega_k \hat A_k$ in place of
$\hat G$ the system simplifies to
\begin{eqnarray*}
\p_t \hat A_i &=& - \hat E_i\,,\\
\p_t \hat E_i &=& \omega^2 \hat A_i + i\omega_i \hat \Gamma\,,\\
\p_t \hat \Gamma &=& 0\,.
\end{eqnarray*}
The eigenvalues and characteristic variables of the symbol are 
\begin{eqnarray*}
0, &\qquad & \hat w^{(0)} = \hat \Gamma\,,\\
\pm i \omega, & \qquad & \hat w^{(\pm)}_i = \hat E_i \mp i \omega \hat
A_i \pm \hat \omega_i \hat \Gamma\,,
\end{eqnarray*}
where $\hat \omega_i = \omega_i/\omega$ and $\omega^2 = \sum_{k=1}^3
\omega_k^2$.  Note that the eigenvalues of the symbol are independent
of the parameter $r$.  To construct a conserved energy we take the
combination
\[
E_C = \frac{1}{2}| \hat w_i^{(+)}|^2 + \frac{1}{2}| \hat w_i^{(-)}|^2 + a |\hat w^{(0)}|^2.
\]
To keep the notation compact we omit the sums.  We need to check that
this conserved quantity is equivalent to\footnote{From the results in
Section \ref{Sec:Stability12} we only need to show that $\hat H$ is
equivalent to $I_{\omega}$, see inequality (\ref{eqn:redScondition}),
which in this case means that there is no $|\hat{A}_i|^2$ term.}
\[
|\hat{\bm{u}}|^2 = |\hat E_i|^2 + \omega^2 |\hat A_i|^2 + |\hat G|^2.
\]

Since
\[
E_C = |\hat E_i|^2 + (1+a)|\hat\Gamma|^2 + \omega^2 |\hat A_i|^2 - 2\Re\left(i\omega_i \hat A_i
\overline{\hat \Gamma} \right),
\]
we get 
\begin{eqnarray*}
&&|\hat E_i|^2 + (1+a -\varepsilon_1) |\hat \Gamma|^2 + \left(1 -
 \frac{1}{\varepsilon_1}\right)\omega^2 |\hat A_i|^2 \le E_C \\
&& \le |\hat E_i|^2 +
 (1+a +\varepsilon_2) |\hat \Gamma|^2 + \left(1 +
 \frac{1}{\varepsilon_2}\right)\omega^2 |\hat A_i|^2,
\end{eqnarray*}
where we used the inequality $\pm 2 \Re(z_1\bar z_2) \le \varepsilon
|z_1|^2 + \varepsilon^{-1} |z_2|^2$ for $\varepsilon>0$.
Choosing $a = 3/2$, $\varepsilon_1 = \varepsilon_2^{-1} = 2$
gives
\[
K_1^{-1} |\hat{\bm{u}}|_{\Gamma}^2 \le E_C \le K_1 |\hat{\bm{u}}|_{\Gamma}^2,
\]
with $K_1 = 3$, where $|\hat{\bm{u}}|_{\Gamma}^2 = |\hat E_i|^2 + \omega^2 |\hat A_i|^2 +
|\hat \Gamma|^2$.  Using the inequality 
\begin{eqnarray}
&&(1-\varepsilon)|z_1|^2 + (1-\varepsilon^{-1})|z_2|^2 \le |z_1+z_2|^2  \label{Eq:z1z2ineq}\\
  &&\le (1+\varepsilon) |z_1|^2 + (1+\varepsilon^{-1}) |z_2|^2, \nonumber
\end{eqnarray}
with $\varepsilon >0$, we have that for any $r$,
$|\hat{\bm{u}}|_{\Gamma}^2$ is equivalent to $|\hat{\bm{u}}|^2$,
i.e. $K_{2}^{-1} |\hat{\bm{u}}|_{\Gamma}^2 \le |\hat{\bm{u}}|^2 \le
K_2 |\hat{\bm{u}}|^2_{\Gamma}$.  We have the uniform estimate in
Fourier space
\begin{eqnarray}
&&|\hat{\bm{u}}(t)|^2 \le K_2 |\hat{\bm{u}}(t)|_{\Gamma}^2 \le K_1K_2 E_C(t) = K_1K_2E_C(0)
  \nonumber\\ 
&&\le K_1^2K_2 |\hat{\bm{u}}(0)|^2_{\Gamma} \le K_1^2K_2^2
  |\hat{\bm{u}}(0)|^2, \label{Eq:kwbEstimate} 
\end{eqnarray}
which implies the estimate in physical space with respect to the norm
\begin{eqnarray}
\| \bm{u}\|^2 = \| A_i \|^2 + \| E_i \|^2 + \| \p_k A_i\|^2 + \| G\|^2,
\end{eqnarray}
with no restrictions on the parameter $r$.

\subsubsection{Discrete analysis}
\label{Sec:discrKWB}

Consider now the semi-discrete system
\begin{eqnarray}
\p_t A_i &=& - E_i\,,\label{Eq:dKWB1}\\
\p_t E_i &=& - D_{+k}D_{-k} A_i + r D^{(2)}_{ik} A_k + D_{0i} G\,,\label{Eq:dKWB2}\\
\p_t G &=& r D_{0k}E_k\,,\label{Eq:dKWB3}
\end{eqnarray}
where $D^{(2)}_{ik}$ is the standard second order accurate
approximation of the second partial derivative.  The procedure is
similar to that at the continuum. We Fourier transform and replace the
variable $\hat G$ with $\hat \Gamma = \hat G + r \frac{i}{h} \sin
\xi_k \hat A_k$ and obtain
\begin{eqnarray*}
\p_t \hat A_i &=& - \hat E_i\,,\\
\p_t \hat E_i &=& \frac{4}{h^2} \Theta_i^2(\xi) \hat A_i + \frac{i}{h} \sin\xi_i
\hat \Gamma\,,\\
\p_t \hat \Gamma &=& 0\,,
\end{eqnarray*}
where $\Theta^2_i(\xi) = \sum_{k=1}^3 \sin^2 \frac{\xi_k}{2} - r
\sin^4\frac{\xi_i}{2}$. 

The eigenvalues of the matrix $k\hat{P}(\xi)$ and the corresponding
characteristic variables are
\begin{eqnarray*}
0, & \qquad & \hat w^{(0)} = \hat \Gamma\,,\\
\pm 2 i \Theta_i(\xi) \lambda, & \qquad & \hat w^{(\pm)}_i = \hat E_i \mp
\frac{2i}{h} \Theta_i(\xi) \hat A_i \pm s_i(\xi)\hat \Gamma\,,
\end{eqnarray*}
where $2s_i\Theta_i = \sin \xi_i$.  The requirement that $\sigma(k
\hat P) \le \alpha_0$ imposes the restriction $r \le 1$ on the
parameter.  If this condition is violated, then the semi-discrete
scheme is unstable (and the fully discrete scheme would be
unconditionally unstable).  Furthermore, for $r = 1$, which
corresponds to the Z1 system, the matrix $\hat{P}(\pm \pi, 0, 0)$
(corresponding to the highest frequency in the $x$ direction) is not
diagonalizable and one can show that the system admits frequency
dependent linearly growing solutions which violate the discrete energy
estimate.

Assume $r<1$.  The expression
\begin{eqnarray*}
E_C &=& \frac{1}{2} |\hat w^{(+)}_i|^2 + \frac{1}{2} | \hat w^{(-)}_i|^2 + a
|\hat \Gamma|^2 \\
&=& |\hat E_i|^2 + (a + s_i^2) |\hat \Gamma|^2 + \frac{4}{h^2} \Theta_i^2
|\hat A_i|^2 - 2\Re\left(
\frac{i}{h}\sin\xi_i \hat A_i \overline{\hat \Gamma}  \right)
\end{eqnarray*}
is conserved.  We want to show that it is equivalent to $|\hat{\bm{u}}|^2 =
|\hat E_i|^2 + \Omega^2 |\hat A_i|^2 + |\hat G|^2$.

We first show that $E_C$ is equivalent to $|\hat{\bm{u}}|_{\Gamma}^2 =
|\hat E_i|^2 + \Omega^2 |\hat A_i|^2 + |\hat \Gamma|^2$.  We
distinguish now between two possibilities: $r \le 0$ and $0<r<1$.  In
either case we have that $| s_i | \le 1$.  In the first case, using
the inequality $\chi_2^2 \le \Theta_i^2 \le (1-r) \chi_2^2$ we get
\begin{eqnarray*}
&&|\hat E_i|^2 + (a - \varepsilon_1) |\hat \Gamma|^2 + \left( 1 -
 \frac{1}{\varepsilon_1}\right) \chi_2^2 |\hat A_i|^2 \le E_C \\
&&\le |\hat E_i|^2 +
 (a + 1 + \varepsilon_2) |\hat \Gamma|^2  + \frac{4}{h^2} \left( 1 -
 r + \frac{1}{\varepsilon_2}\right) \chi_2^2 |\hat A_i|^2.
\end{eqnarray*}
If we take, for example, $a \ge 3$, $\varepsilon_1 = 2$,
$\varepsilon_2 = 1/2$, then there exist constants $K_1$ and $K_2$
such that $K_1 |\hat{\bm{u}}|_{\Gamma}^2 \le E_C \le K_2
|\hat{\bm{u}}|_{\Gamma}^2$. 

For the case $0< r < 1$, using the inequality $(1-r) \chi_2^2 \le
\Theta_i^2 \le \chi_2^2$ we get
\begin{eqnarray*}
&&|\hat E_i|^2 + (a-\varepsilon_1) |\hat \Gamma|^2 + \left(1
 -r -\frac{1}{\varepsilon_1}\right) \chi_2^2 |\hat A_i|^2 \le E_C \le\\
&&|\hat E_i|^2  +
 (a + 1 + \varepsilon_2) |\hat \Gamma|^2  + \frac{4}{h^2} \left( 1 + \frac{1}{\varepsilon_2}\right) \chi_2^2 |\hat A_i|^2.
\end{eqnarray*}
If we choose $a>\varepsilon_1 > 1/(1-r)$ we have the equivalence to
$|\hat{\bm{u}}|_{\Gamma}^2$.  On the other hand, using
\begin{equation}
\frac{1}{h}|\sin\xi_k| \le |\Omega|\,, \label{Eq:Omegaineq}
\end{equation}
one can show that the norms $|\hat{\bm{u}}|_{\Gamma}^2$ and
  $|\hat{\bm{u}}|^2$ are equivalent.  This proves stability with
  respect to the norm
\begin{eqnarray}
\left( \| A_i \|_h^2 + \| E_i \|_h^2 + \| D_{+k}A_i\|_h^2 + \| G
\|_h^2 \right)^{1/2}.
\end{eqnarray}

Note that the Cauchy problem for the continuum system is well-posed
for all values of $r$, but the discrete system is stable only for
$r<1$.  For $r \le 1/2$ the von Neumann condition gives a Courant
limit of $\lambda \le \alpha_0 / (2\sqrt{3-r})$.  Moreover, the
numerical speeds of propagation depend on $r$.

\subsection{The Nagy-Ortiz-Reula system}

The NOR formulation of Einstein's equations linearized about Minkowski
space with zero shift and densitized lapse ($\alpha =
\det(\gamma_{ij})^{1/2}$) has the form
\begin{eqnarray}
\p_t \gamma_{ij} &=& -2K_{ij}\,,\label{Eq:NOR1}\\
\p_t K_{ij} &=& -\frac{1}{2} \p^k\p_k \gamma_{ij} + \frac{r}{2} \p_i
\p_j \tau + \p_{(i} f_{j)}\,,\label{Eq:NOR2}\\
\p_t f_i &=& r \p_i K\,,\label{Eq:NOR3}
\end{eqnarray}
where $\tau = \delta^{kl}\gamma_{kl}$.
This system corresponds to the one in \cite{GG2} with the choice of
parameters $a=b=\sigma=1$, $c=0$ and $\rho = r+2$.  It is obtained
from the ADM system with densitized lapse by introducing the variables
$f_i = \p_j \gamma_{ij} - \p_i \tau$, which are used in the
evolution equations for the $K_{ij}$ variables, and adding the
momentum constraint to the time derivative of the new variables.

\subsubsection{Continuum analysis}

We Fourier transform the system and introduce $\hat \Gamma_i = \hat
f_i + \frac{r}{2} i \omega_i \hat \tau$, obtaining
\begin{eqnarray*}
\p_t \hat \gamma_{ij} &=& -2\hat K_{ij}\,,\\
\p_t \hat K_{ij} &=& \frac{1}{2}\omega^2 \hat \gamma_{ij} + i \omega_{(i} \hat \Gamma_{j)}\,,\\
\p_t \hat \Gamma_i &=& 0\,.
\end{eqnarray*}

The eigenvalues and characteristic variables associated with the
symbol are
\begin{eqnarray*}
0,&\qquad& \hat w^{(0)}_i = \hat \Gamma_i\,,\\
\pm i \omega\,, &\qquad & \hat w^{(\pm)}_{ij} = \hat K_{ij} \mp \frac{1}{2} i \omega
\hat \gamma_{ij} \pm \hat \omega_{(i} \hat \Gamma_{j)}\,.
\end{eqnarray*}

Proceeding in the usual manner we construct a conserved quantity and
show that it is equivalent to 
\[
|\hat{\bm{u}}|^2 = |\hat K_{ij}|^2 + \omega^2 |\hat \gamma_{ij}|^2 + |\hat f_i|^2.
\]
We have
\begin{eqnarray*}
E_C &=& \frac{1}{2} |\hat w^{(+)}_{ij}|^2 +  \frac{1}{2} |\hat w^{(-)}_{ij}|^2 +
a | \hat w^{(0)}_i|^2 \\
&=& |\hat K_{ij}|^2 + |\hat \omega_{(i}\hat \Gamma_{j)}|^2 + \frac{1}{4}\omega^2 |
\hat \gamma_{ij}|^2 - \Re\left(i \omega_i \hat \gamma_{ij} \overline{
  \hat \Gamma_j}\right) + a | \hat \Gamma_j|^2 .
\end{eqnarray*}

Since 
\begin{eqnarray*}
&&0 \le | \hat\omega_{(i}\hat \Gamma_{j)}|^2 \le | \hat\omega_i\hat \Gamma_j|^2
\le |\hat \Gamma_i|^2 -\frac{\omega^2}{\varepsilon_1} |\hat \gamma_{ij}|^2 - \varepsilon_1 |
\hat \Gamma_i|^2 \le -2\Re\left(i \omega_i \hat \gamma_{ij}
\overline{\hat \Gamma_j}\right) \\
&&\le
\frac{\omega^2}{\varepsilon_2} | \hat \gamma_{ij}|^2 + \varepsilon_2
|\hat \Gamma_i|^2,
\end{eqnarray*}
we obtain the equivalence  with $|\hat{\bm{u}}|^2_{\Gamma}$,
\begin{eqnarray*}
&&|\hat K_{ij}|^2 + \frac{1}{4}\left(1-\frac{1}{\varepsilon_1}\right) \omega^2
 |\hat \gamma_{ij}|^2 + (a-\varepsilon_1)|\hat \Gamma_i|^2 \le E_C\\
&& \le |\hat K_{ij}|^2 + \frac{1}{4}\left(1+\frac{1}{\varepsilon_2}\right) \omega^2
 |\hat \gamma_{ij}|^2 + (1+a+\varepsilon_2)|\hat \Gamma_i|^2,
\end{eqnarray*}
by choosing $a=3$, $\varepsilon_1=2$, $\varepsilon_2=1$. Finally,
noting that $|\hat \tau|^2 \le 3 |\hat \gamma_{ij}|^2$ one can show that
$|\hat{\bm{u}} |_{\Gamma}^2$ and $|\hat{\bm{u}}|^2$ are equivalent.

\subsubsection{Discrete analysis}

We consider the standard second order accurate discretization of system
(\ref{Eq:NOR1})--(\ref{Eq:NOR3}). The semi-discrete system is
\begin{eqnarray}
\p_t \gamma_{ij} &=& -2K_{ij}\,,\\
\p_t K_{ij} &=& -\frac{1}{2}D_{+k}D_{-k} \gamma_{ij} + \frac{r}{2}
D^{(2)}_{ij} \tau + D_{0(i} f_{j)}\,,\\
\p_t f_i &=& rD_{0i}K\,.
\end{eqnarray}

Taking the Fourier transform and introducing $\hat \Gamma_i = \hat f_i
+ \frac{r}{2} \frac{i}{h} \sin \xi_i \hat \tau$ gives
\begin{eqnarray*}
\p_t \hat \gamma_{ij} &=& -2\hat K_{ij}\,,\\
\p_t \hat K_{ij} &=& \frac{1}{2} \Omega^2 \hat \gamma_{ij} +
\frac{r}{2} \hat\Delta_{ij} \hat \tau + \frac{i}{h} \sin\xi_{(i}
\hat \Gamma_{j)}\,,\\
\p_t \hat \Gamma_i &=& 0\,,
\end{eqnarray*}
where 
\begin{eqnarray*}
\hat \Delta_{ij} = \left\{ \begin{array}{l@{\quad}c}
0 & i \neq j\\
-\displaystyle\frac{4}{h^2} \sin^4\frac{\xi_i}{2}& i = j
\end{array}\right..
\end{eqnarray*}

The eigenvalues of $k\hat P$ and the corresponding characteristic
variables are 
\begin{eqnarray*}
0, &\qquad& \hat w^{(0)}_i = \hat \Gamma_i\,,\\
\pm 2 i \Theta \lambda, & \qquad & \hat w^{(\pm)} = \hat K \mp \frac{i}{h} \Theta
\hat \tau \pm \frac{\sin\xi_i}{2\Theta} \hat \Gamma_i\,,\\
\pm 2i\chi_2\lambda,  &\qquad& \hat w^{(\pm)}_{ij} = \hat K_{ij} \mp \frac{1}{2} i\Omega
\hat \gamma_{ij} \pm \frac{\sin\xi_{(i}\hat \Gamma_{j)}}{2\chi_2}\,, \quad i\neq j\,,\\
&& \hat w^{(\pm)}_i = \left(\tilde K_{ii} \mp \frac{1}{2}i\Omega
\tilde \gamma_{ii} \pm \frac{\sin\xi_{i}\tilde \Gamma_{i}}{2\chi_2}\right)^{\rm TF},
\end{eqnarray*}
where $\Theta^2 = \chi_2^2 - r \sum_{k=1}^3 \sigma_i^4$,
$\sigma_i^4 = \sin^4\frac{\xi_i}{2}$, $\sigma_i^4 \tilde K_{ii} =
\hat K_{ii}$, $\sigma_i^4 \tilde \gamma_{ii} = \hat \gamma_{ii}$, 
$\sigma_i^4 \tilde \Gamma_i = \hat \Gamma_i$, and $A^{\rm TF}_{ij} =
(A_{ij} - \delta_{ij} A/3)$.

\begin{figure}[ht]
\begin{center}
\includegraphics[width=8cm]{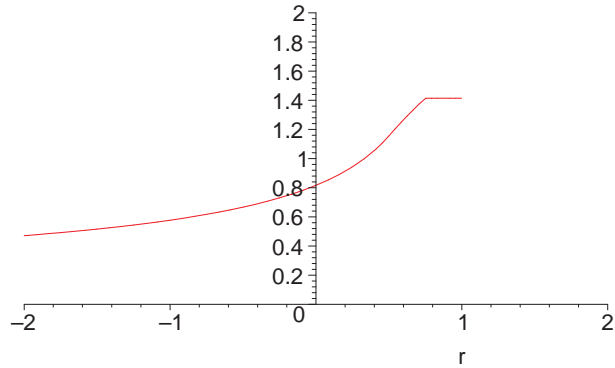}
\end{center}
\caption{The von Neumann condition for the second order accurate
  discretization of the NOR system in 3D using 4RK as a function of
  the parameter $r$.  For $r>1$ the scheme is unconditionally
  unstable.}
\label{Fig:courantnor}
\end{figure}

Note that stability demands that $r<1$ ($\rho<3$).  Furthermore, the
von Neumann condition depends on the value of this parameter.
Explicitly, this is
\[
\lambda \le \frac{\alpha_0}{2\max_{|\xi_i|\le \pi}\{\Theta,\chi_2\}}
\]
and its dependence on $r$ is illustrated in Figure
\ref{Fig:courantnor}.  This is in contrast to the fact that at the
continuum $r$ has no influence on the characteristic speeds or the
hyperbolicity of the system.

We now restrict ourselves to the case $r=0$ and prove numerical
stability.  In this case the characteristic variables associated with
the non trivial eigenvalues are
\begin{equation}
\hat w^{(\pm)}_{ij} = \hat K_{ij} \mp \frac{1}{2}i\Omega
\hat \gamma_{ij} \pm \frac{\sin\xi_{(i}\hat \Gamma_{j)}}{2\chi_2}\,.
\end{equation}
A conserved quantity is 
\begin{eqnarray*}
E_C &=& \frac{1}{2} | \hat w^{(+)}_{ij}|^2 + \frac{1}{2} | \hat w^{(-)}_{ij}|^2 +
a | \hat w_i^{(0)}|^2\\
&=&|\hat K_{ij}|^2 + |s_{(i}\hat \Gamma_{j)}|^2 + \frac{\Omega^2}{4} |
\hat \gamma_{ij}|^2- \Re\left(\frac{i}{h} \sin \xi_i \hat \gamma_{ij}
\overline{\hat \Gamma_j}\right)  + a |\hat \Gamma_i|^2,
\end{eqnarray*}
where $2\chi_2 s_i = \sin\xi_i$.  

Since
\begin{eqnarray*}
&&|s_i| \le 1\\
&&0 \le | s_{(i}\hat \Gamma_{j)}|^2 \le | s_i \hat \Gamma_j |^2 \le |
\hat \Gamma_i|^2-\frac{4}{\varepsilon_1 h^2} \chi_2^2 |\hat \gamma_{ij}|^2 - \varepsilon_1
| \hat \Gamma_i|^2 \\
&&\le - 2\Re \left(\frac{i}{h} \sin \xi_i \hat \gamma_{ij}
\overline{\hat \Gamma_j}\right)\le \frac{4}{\varepsilon_2 h^2} \chi_2^2 |\hat \gamma_{ij}|^2 +
\varepsilon_2 | \hat \Gamma_i|^2, 
\end{eqnarray*}
we have the equivalence with $|\hat{\bm{u}}|_{\Gamma}^2$.  Inequality
(\ref{Eq:Omegaineq}) guarantees the equivalence of the latter with
$|\hat{\bm{u}}|^2$.  This completes the proof of stability with respect to
the norm
\begin{equation}
\left( \|\gamma_{ij}\|_h^2 + \|K_{ij}\|_h^2 +
\|D_{+k}\gamma_{ij}\|_h^2 + \|f_i\|_h^2
\right)^{1/2}.
\end{equation}

\subsection{The ADM system}

With a densitized lapse function, $\alpha = \det(\gamma_{ij})^{1/2}$,
the ADM equations linearized about the Minkowski solution in Cartesian
coordinates take the form
\begin{eqnarray}
\p_t \gamma_{ij} &=& -2K_{ij}\,,\label{Eq:ADMdens1}\\
\p_t K_{ij} &=& \p_k\p_{(i}\gamma_{j)k} - \frac{1}{2} \p^k\p_k
\gamma_{ij} -\p_i\p_j \tau\,.\label{Eq:ADMdens2}
\end{eqnarray}
The symbol $\hat P(i\omega)$ of
(\ref{Eq:ADMdens1})--(\ref{Eq:ADMdens2}) is not diagonalizable and
neither is that of its differential nor its pseudo-differential
reduction.  The family of solutions in which the only non vanishing
components are $\gamma_{1A} = \sin(\omega x)t$, $K_{1A} = -\sin(\omega
x)/2$, where $A=2,3$, can be used to explicitly show instability.  It
gives
\begin{equation}
\frac{\| \bm{u}(t,\cdot) \|}{\| \bm{u}(0,\cdot)\|} = \left( 1+ 4t^2 + 4\omega^2
  t^2\right)^{1/2},
\end{equation}
where $\| u(t,\cdot) \|^2 = \| \gamma_{ij}(t,\cdot) \|^2 + \|
K_{ij}(t,\cdot) \| + \| \p_k \gamma_{ij}(t,\cdot) \|^2$.  The ratio
cannot be bounded by $K e^{\alpha t}$ with $K$ and $\alpha$
independent of $\omega$.

To see that the second order accurate standard discretization is
unstable we take $\gamma_{1A} = (-1)^jt$ and $K_{1A} = (-1)^{j+1}/2$.
As in the continuum, the ratio
\begin{equation}
\frac{\| \bm{v}(t)\|_{h,D_+}}{\|\bm{v}(0)\|_{h,D_+}} = \left( 1+ 4t^2 +
  16\frac{t^2}{h^2}\right)^{1/2} \label{Eq:ratioADM}
\end{equation}
cannot be bounded.  We can nevertheless compute the von Neumann
condition, which is given by
\begin{equation}
\lambda \le \frac{\sqrt{3} \alpha_0}{2\sqrt{7d}}. \label{Eq:CL_ADM}
\end{equation}

In \cite{Mexico1} stability tests were done with the non linear
version of this formulation.  The domain used consisted of a thin
channel, with an even number $N$ of grid points in one spatial
direction and 3 grid points in the other two directions.  By taking
this into account we see that modes corresponding to the frequencies
$\xi_1 = \pi$, and $\xi_2= \xi_3 = 2\pi/3$ grow exponentially if
$\lambda > 0.4163$.  Figure 2 in \cite{Mexico1} confirms that with a
Courant factor of $\lambda = 0.5$ there is a violation of
the von Neumann condition.\footnote{A one-dimensional von Neumann
analysis gives the limit (\ref{Eq:CL_ADM}) with $d=1$ and $\alpha_0 =
2$, which corresponds to $0.655$.  However, this would not capture the
fact that there could be exponentially growing modes with non trivial
dependence in the two thin directions.}

Although the symbol associated with the continuum system
(\ref{Eq:ADMdens1}) and (\ref{Eq:ADMdens2}) has four Jordan blocks of
size two for any $\omega$, interestingly, the symbol associated with
the semi-discrete problem obtained with the standard second order
accurate discretization can have rather different properties.  For
Fourier modes traveling in directions parallel to the axis the
continuum result still holds.  However, for Fourier modes not parallel
to any of the axis, we found that the symbol may have fewer Jordan
blocks.  For some Fourier frequencies we even noticed that the symbol
is diagonalizable.  There is no conflict between this observation and
the fact that the continuum problem is ill-posed.  As shown at the
beginning of this subsection the discrete initial value problem for
the ADM system is also ill-posed.  In the limit of high resolution,
$h\to 0$ ($\xi\to 0$ and $\omega$ fixed), the discrete symbol is a
perturbation of the continuum one\footnote{Note that in general by
perturbing a non diagonalizable matrix one obtains a diagonalizable
matrix, so the diagonalizability of the discrete ADM symbol for some
frequencies should not be so surprising.}
\[
\hat P_d = \hat P_c + O(h^2).
\]
Even though for some frequencies $\hat P_d$ is diagonalizable, the
characteristic variables become degenerate in the limit $h\to 0$,
which implies that the discrete symmetrizer becomes unbounded (it is
not possible to find a $K$, independent of $h$, satisfying inequality
(\ref{Eq:dequiv})).

\subsection{The Z4 system}
\label{Sec:Z4}

The same family of solutions that was used to show instability of the
discretized ADM equations can be used for the standard discretization of
the linearized $Z4$ system \cite{BLPZ1}
\begin{eqnarray*}
\p_t \alpha &=& -f(K-m\Theta)\,,\\
\p_t \gamma_{ij} &=& -2K_{ij}\,,\\
\p_t K_{ij} &=& -\p_i\p_j \alpha  - \frac{1}{2} \p_k\p_k\gamma_{ij} +
\p_{k}\p_{(i}\gamma_{j)k} - \frac{1}{2} \p_i\p_j \tau + 2 \p_{(i}
Z_{j)}\,,\\
\p_t \Theta &=& \frac{1}{2}( \p_k\p_l \gamma_{kl} - \p_k\p_k \tau) + \p_kZ_k\,,\\
\p_t Z_i &=&  \p_k K_{ik} - \p_i K + \p_i \Theta\,,
\end{eqnarray*}
for any values of the parameters $f$ and $m$.  This instability,
however, is not present if the $D_0^2$ discretization is used as in
\cite{Bonetal}, in conjunction with the $D_0$-norm.  Furthermore, it
is possible that artificial dissipation may cure this instability of
the standard discretization, at least for $0<f\neq 1$ or $1=f=m/2$,
since in this case the continuum Cauchy problem is well-posed.  Note
that while we use the same family of solutions that was used to show
instability for the ADM case, the two cases are very different: While
the ADM instability is due to the lack of well-posedness of the
continuum equations, the problem with the Z4 system arises purely at
the discrete level, and can be traced back to the difference in
structure between the principal symbols of the pseudodifferential
first order reductions of the continuum and discrete equations, see
Eqs.  (\ref{eqn:PRprimed}) and (\ref{eqn:PR}). For second order in
space systems diagonalizability of the discrete symbol is not implied
by diagonalizability of the continuum symbol.

The ADM and Z4 examples suggest a simple criterion that can be used to
rule out certain schemes.  Any first order in time, second order in
space system of PDEs which gives rise to an ill-posed problem when the
first order and mixed second order spatial derivatives are dropped
will result in an unstable scheme if the standard discretization is
used and no artificial dissipation is added.  This is a consequence of
the fact that grid modes with the highest frequency belong to the
kernel of the $D_0$ operator.  Although the $D_0^2$ discretization
gives stable schemes with respect to the $D_0$-norm, provided that the
continuum problem is well-posed, it suffers from the limitations
described in section \ref{Sec:D02}.

\section{Testing stability}
\label{Sec:Testing}

When dealing with variable coefficient or non linear problems it can
be difficult, if not impossible, to prove stability with respect to a
certain norm.  Numerical experiments are often the only option.  Given
a discretization of the linear initial value problem
(\ref{Eq:Cauchy1}) and (\ref{Eq:Cauchy2}), a stability test should be
aimed at establishing the existence of the constants $\alpha$ and $K$,
independent of the initial data and for all $h \le h_0$ (and possibly
$k\le \lambda_0 h$), by computing the ratio between a suitable
discrete norm at time-step $t_n=nk$ and its initial value,
\begin{equation}
\frac{\| v^n \|}{\| v^0 \|} \le K e^{\alpha t_n}.\label{Eq:stabtest}
\end{equation}
Although it is not possible to infer stability by examining a finite
number of numerical experiments (one would have to explore the entire
set $h \le h_0$ that appears in the definition of stability), it is
usually not difficult to spot a trend of behavior as the resolution is
increased.  To ensure that a wide range of frequencies is excited,
random initial data can be used \cite{RobStability}, as no smoothness
assumptions are used in the definition of stability.

In the examples of first order in time, second order in space
hyperbolic systems for which we are able to determine stability, we
use a norm which is the discrete version of the continuum one.  The
derivatives are approximated using the one-sided operators $D_+$ (or,
equivalently, $D_-$) rather than $D_0$.  For the NOR system, for
example, we use the square root of the expression
\[
\sum_{i,j=1}^3 \| \gamma_{ij} \|_h^2 + \sum_{i,j=1}^3 \|K_{ij}\|_h^2 +
\sum_{k,i,j=1}^3\|D_{+k}\gamma_{ij}\|_h^2 + \sum_{i=1}^3\|f_i\|_h^2.
\]
If, as we vary the initial data and the resolution, the experiments
indicate that the constants $\alpha$ and $K$ in (\ref{Eq:stabtest})
exist, then one would conclude that the scheme appears to be stable.
If not, the scheme appears to be unstable.

In the nonlinear case, if the problem has a sufficiently smooth
solution $u_0$, then to first approximation the error equation can be
linearized about $u_0$ and convergence follows if the linearized
equation is stable (Sec.~5.5 in \cite{GKO-Book}).  Establishing
stability experimentally using the linearized equations would not be
very practical.  However, convergence to a known exact solution can be
tested directly and it avoids many complications.  Rather than testing
for stability, one could test convergence in a more demanding way:
initial data can be chosen which is not smooth, but is accurate to the
correct order in the appropriate norm.  For instance, for the NOR
system, one would use the square root of
\[
\sum_{i,j=1}^3 \| \delta \gamma_{ij} \|_h^2 + \sum_{i,j=1}^3
\|\delta K_{ij}\|_h^2 +
\sum_{k,i,j=1}^3\|D_{+k}(\delta \gamma_{ij})\|_h^2 +
\sum_{i=1}^3\|\delta f_i\|_h^2,
\]
where $\delta v = v - u_0$, and one could add random noise to the
initial data with amplitude $h^p$ for the $K_{ij}$ and $f_i$ variables
and $h^{p+1}$ for the $\gamma_{ij}$ variables.  The scheme is
convergent around the solution $u_0$ if the $D_+$-norm of the error at
time $T$ is of order $h^p$.  In particular, this implies that for a
convergent scheme the discrete $L_2$ norm of the error is of order
$h^p$ if the $D_+$-norm of the initial error is of order $h^p$.

Finally, we note that the notion of {\em robust stability} introduced
in \cite{RobStability} does not imply nor follows from the concept of
numerical stability investigated in this paper.

\subsection{Numerical tests}

We have performed numerical tests to complement the analytical
stability results of Sec.~\ref{Sec:Applications}.  

For each run, the numerical grid has dimensions $50 \rho \times 4
\times 4$, where $\rho = 1,2, 4, 8$ parameterizes the resolution, and
we impose periodic boundary conditions.  The coordinate domain is $x,
y, z \in [-0.5,0.5)$. The time integrator is RK4 with Courant factor
$\lambda = 0.5$.  We choose random noise of order unity as initial
data (except for the Z4 tests, see below) so that many discrete
Fourier modes are present in the initial data.  Empirically, we find
that using smooth initial data in the constant coefficient problems of
this paper can make it difficult to observe an instability.  This was
also noticed in the nonlinear case in \cite{CPST}.

\begin{figure}[ht]
\begin{center}
\includegraphics{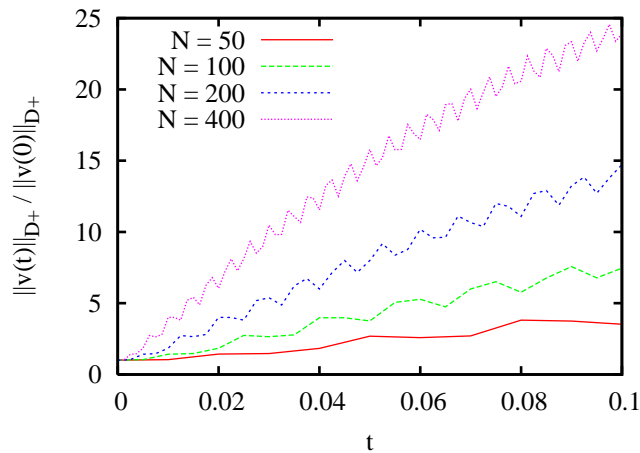}
\end{center}
\caption{Linearized ADM stability test}
\label{Fig:stabtestadm}
\end{figure}

Figure \ref{Fig:stabtestadm} shows the results for the ADM system. The
apparent trend is that as the resolution is increased, and higher
frequency Fourier modes are present in the initial data, the ratio of
the $D_+$-norm of the solution to its initial value at any given time
increases.  It appears that there is no $K, \alpha$ such that this
quantity can be bounded by a function $K e^{\alpha t_n}$, and this
indicates that the system is unstable.

\begin{figure}[ht]
\begin{center}
\includegraphics{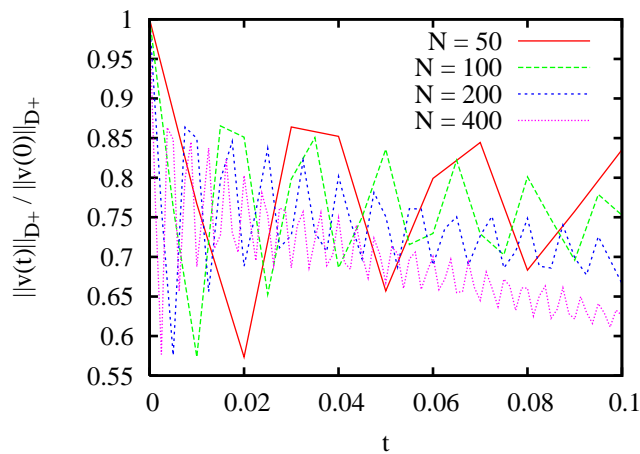}
\end{center}
\caption{Linearized NOR stability test, $r = 0$}
\label{Fig:stabtestnor}
\end{figure}

In Figure \ref{Fig:stabtestnor} we show the results of the stability
test for the linearized NOR system.  The results suggest that the
ratio of the $D_+$-norm of the solution to its initial value remains
bounded, and hence that the system is stable.  This reflects the
analytic result that we proved in Sec.~\ref{Sec:Applications}.

\begin{figure}[ht]
\begin{center}
\includegraphics{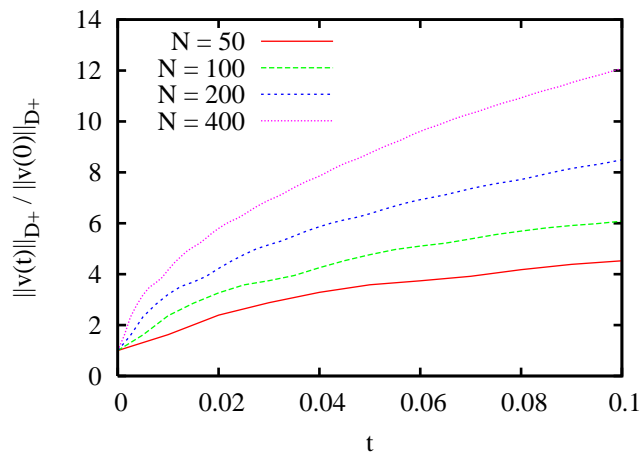}
\end{center}
\caption{Linearized Z4 stability test, $f = 1, m = 2$. Initial data
consists of random values in $K_{22}$ and $K_{33}$, and all other
variables are zero.}
\label{Fig:stabtestz4}
\end{figure}

Showing the instability of the Z4 system was more complicated.  In
this case, it was not sufficient to use random initial data of order
unity in all variables.  When this was attempted, the ratio of the
$D_+$-norm to its initial value remained bounded.  In order to
numerically demonstrate the instability, we used knowledge of the
exact solution that violates the estimate.  Random data of order unity
was given to the variables $K_{22}$ and $K_{33}$ and the remaining
variables were set to zero. The test results for the linearized Z4
system are shown in Figure \ref{Fig:stabtestz4}, and confirm that this
system is unstable.

\begin{figure}[ht]
\begin{center}
\includegraphics{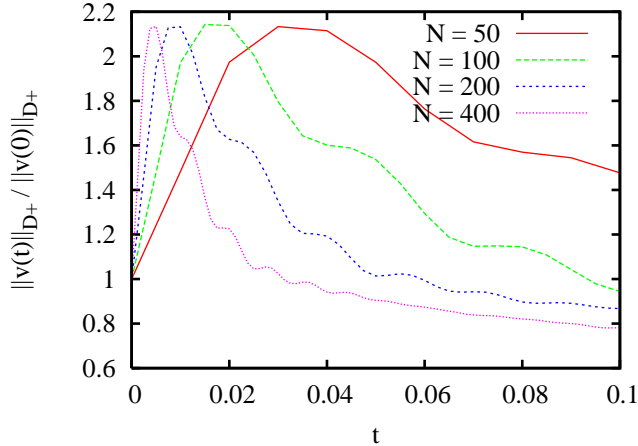}
\end{center}
\caption{Linearized Z4 stability test with dissipation $\sigma =
0.02$, $f = 1, m = 2$. Initial data consists of random values in
$K_{22}$ and $K_{33}$, and all other variables are zero.}
\label{Fig:stabtestz4diss}
\end{figure}

When artificial dissipation with $\sigma = 0.02$ is used, the
linearized Z4 system tested with the same initial data shows no sign
of instability.  See Figure \ref{Fig:stabtestz4diss}.

The example of the Z4 system shows that numerical testing of stability
is not always straightforward, and that schemes which appear stable
for simple test cases may in fact be unstable.   All tests were done
using the standard second order accurate discretization.

\section{Discussion}
\label{Sec:Discussion}

In this work we extended the notion of numerical stability of finite
difference approximations to include hyperbolic systems that are first
order in time and second order in space. We considered the standard
discretization of the wave equation, a generalization of the KWB
formulation of electromagnetism and the NOR formulation of Einstein's
equations linearized about the Minkowski solution.  By analyzing the
symbol of the second order system, and constructing a discrete
symmetrizer, we were able to prove stability in a discrete norm
containing one-sided difference operators, provided that the von
Neumann condition is satisfied.  Consistency and stability with
respect to the $D_+$-norm imply convergence with respect to the
discrete $L_2$ norm.  We also found that in some cases
($r\ge 1$ in the NOR and generalized KWB systems, and Z4) standard
discretizations of well-posed continuum problems can lead to
unconditionally unstable schemes.  This is closely related to the
instability of the fully second order shifted wave equation
investigated in \cite{C2}, but our examples contain no shift terms.

Our analysis of discretizations of first order in time hyperbolic
systems shows that in the first order in space case there is a clear
correspondence between strong hyperbolicity and numerical
stability, and between characteristic speeds and Courant limits.  See
inequality (\ref{Eq:stablimFOSH}) and Eq.~(\ref{Eq:symmfirst}).
In the second order in space case, on the other hand, the mixing of
$D_{\pm}$ and $D_0$ operators breaks this correspondence.  
To restore the correspondence one could use the $D_0^2$
discretization, however, as discussed in Sec.~\ref{Sec:D02}, this
can lead to difficulties.

In Sec.~\ref{Sec:Z4} we propose a simple criterion that can be used to
rule out certain schemes when the standard discretization is used and
no artificial dissipation is added.  This criterion detects schemes in
which the highest frequency mode grows faster as the resolution is
increased.

We also discuss stability tests for second order in space systems.
These tests should be aimed at establishing the existence, for
sufficiently small $h$, of the constants $K$ and $\alpha$ that appear
in the definition of stability with respect to the $D_+$-norm.  In the
nonlinear case the situation is more complicated.  In this case we
suggest, when an exact smooth solution of the continuum problem is
available, to do convergence tests with initial data given by that of
the continuum problem plus random noise of order $h^p$ with respect to
the $D_+$-norm (see Sec.~\ref{Sec:Testing}).

Although our analysis was restricted to the constant coefficient case,
we expect that for the variable coefficient case generalizations of
results similar to those presented in Sec.~6.6 of \cite{GKO-Book}
for first order hyperbolic systems, where artificial dissipation plays
an important role, might apply.


\section{Acknowledgments}

We wish to thank Carsten Gundlach and Olivier Sarbach for helpful
discussions and suggestions.  This research was supported by a Marie
Curie Intra-European Fellowship within the 6th European Community
Framework Program.



\appendix 

\section{Time integrators}
\label{Sec:Timeintegrators}

In this work we restrict our attention to the following three time
integrators: 3rd and 4th order Runge-Kutta, and iterative
Crank-Nicholson \cite{T}.  Given a system of ordinary differential
equations, $dy/dt = f(t,y(t))$, these integrators are defined as
\begin{itemize}
\item[3RK]
\begin{eqnarray*}
k_1 &=& k f(t_n,y^n)\\
k_2 &=& k f(t_n+k/2,y^n+k_1/2)\\
k_3 &=& k f(t_n+3k/4,y^n+3k_2/4)\\
y^{n+1} &=& y^n + (2k_1+3k_2+4k_3)/9
\end{eqnarray*}
\item[4RK]
\begin{eqnarray*}
k_1 &=& k f(t_n,y^n)\\
k_2 &=& k f(t_n+k/2,y^n+k_1/2)\\
k_3 &=& k f(t_n+k/2,y^n+k_2/2)\\
k_4 &=& k f(t_n+k,y^n+k_3)\\
y^{n+1} &=& y^n + (k_1+2k_2+2k_3+k_4)/6
\end{eqnarray*}
\item[ICN]
\begin{eqnarray*}
k_1 &=& k f(t_n,y^n)\\
k_2 &=& k f(t_n+k/2,y^n+k_1/2)\\
k_3 &=& k f(t_n+k/2,y^n+k_2/2)\\
y^{n+1} &=& y^n + k_3
\end{eqnarray*}
\end{itemize}

\section{Some numerical properties of first and second order systems}
\label{Sec:NumProp12}

In this section we assume that the time integrator is one of those
discussed in Appendix \ref{Sec:Timeintegrators}.  We consider standard
second and fourth order accurate discretizations of the following two
toy model problems
\begin{equation}
u_t = u_x\,, \label{Eq:advective}
\end{equation}
and
\begin{equation}
\phi_t = \Pi\,, \qquad \Pi_t = \phi_{xx}\,. \label{Eq:wave}
\end{equation}

\begin{figure}[ht]
\begin{center}
\includegraphics[width=5cm]{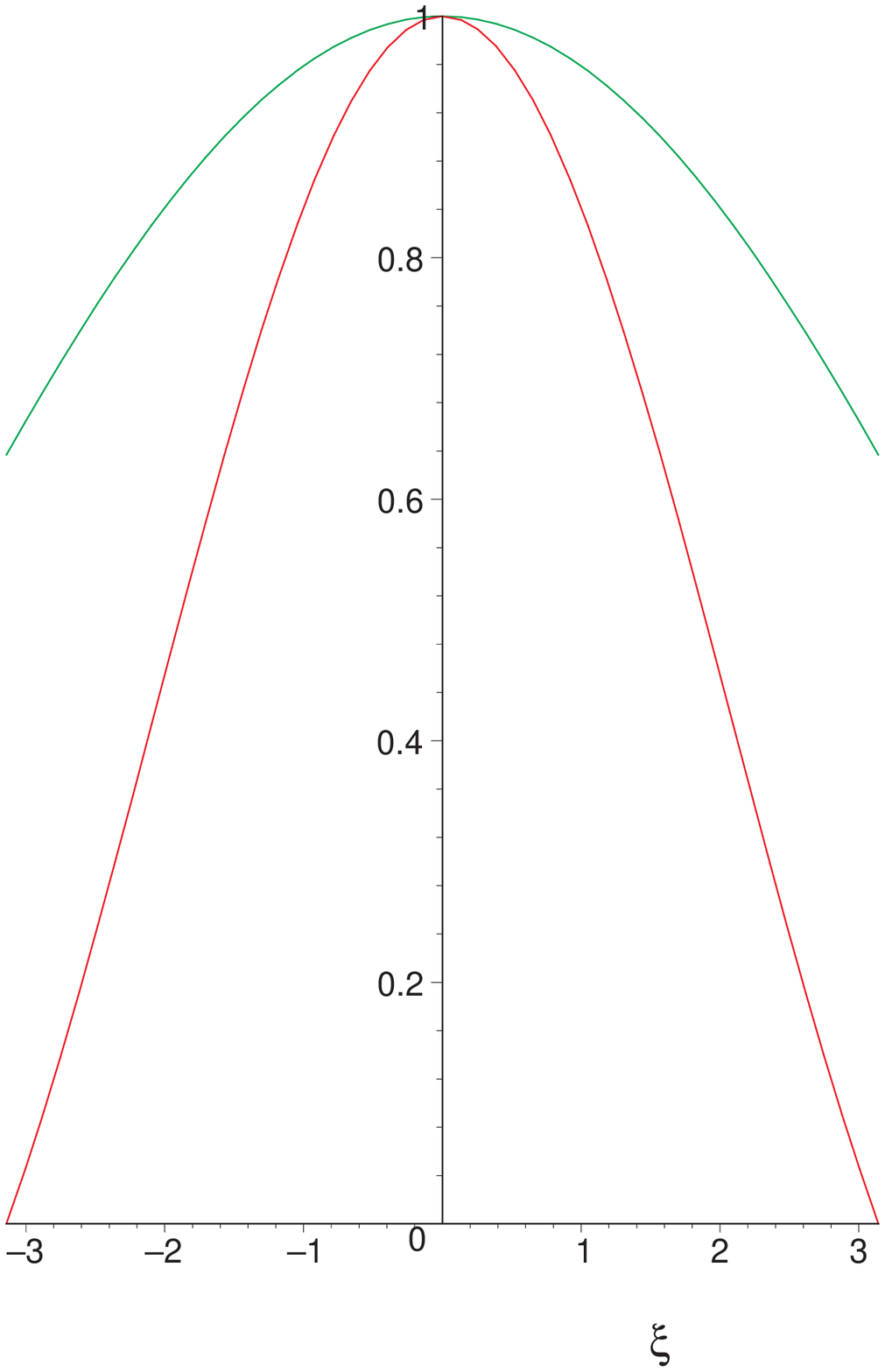}
\includegraphics[width=5cm]{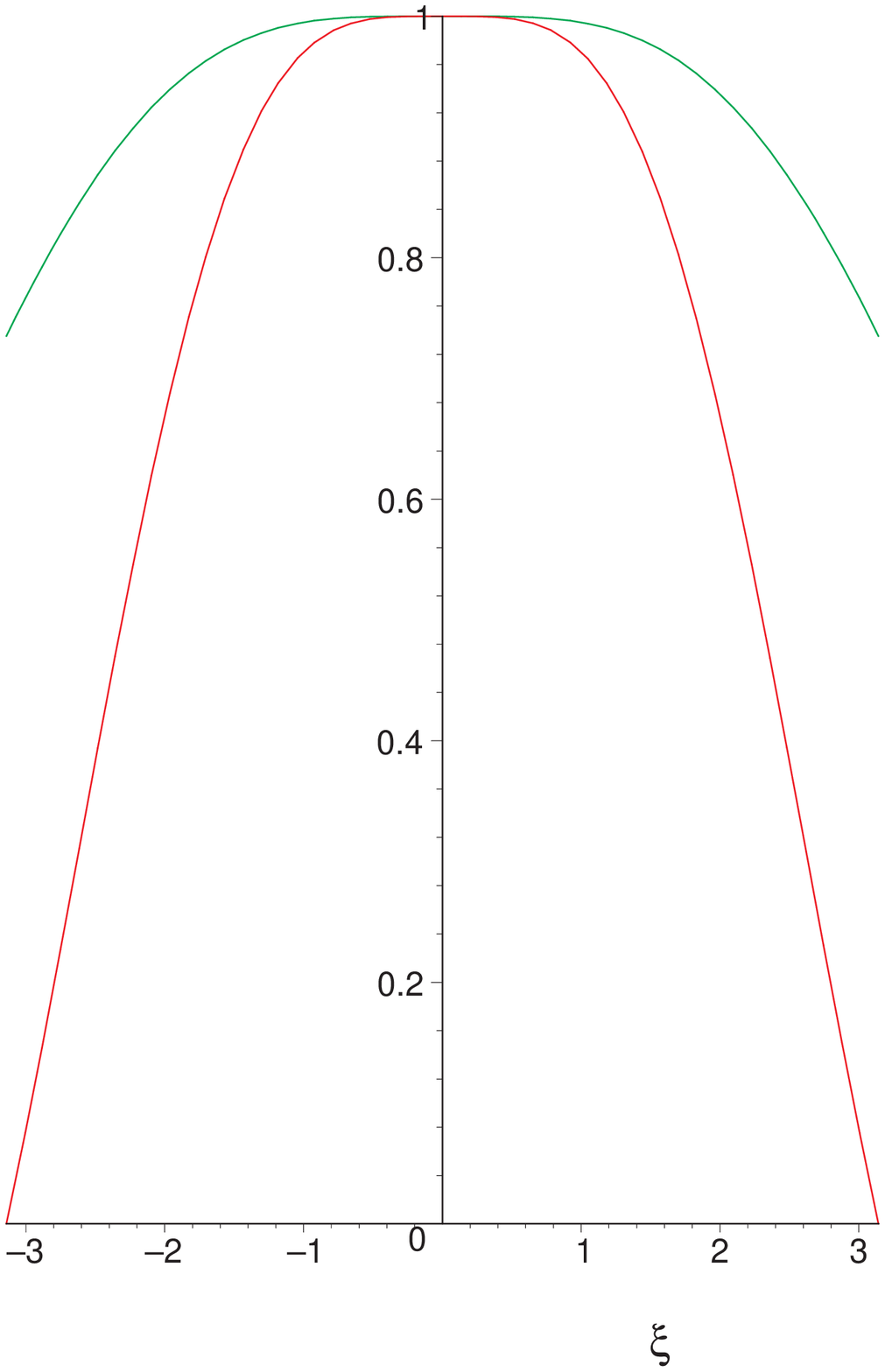}\\
\includegraphics[width=5cm]{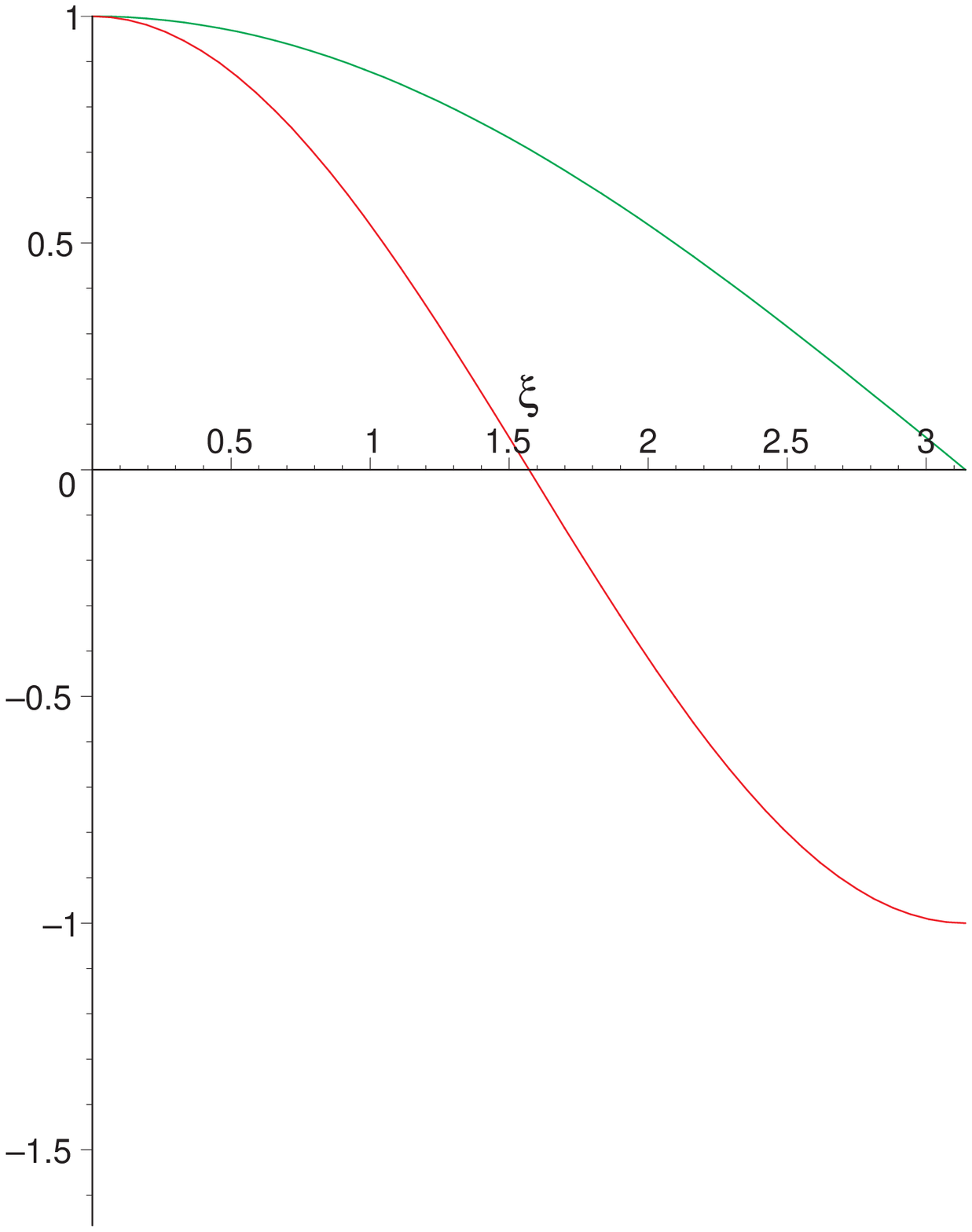}
\includegraphics[width=5cm]{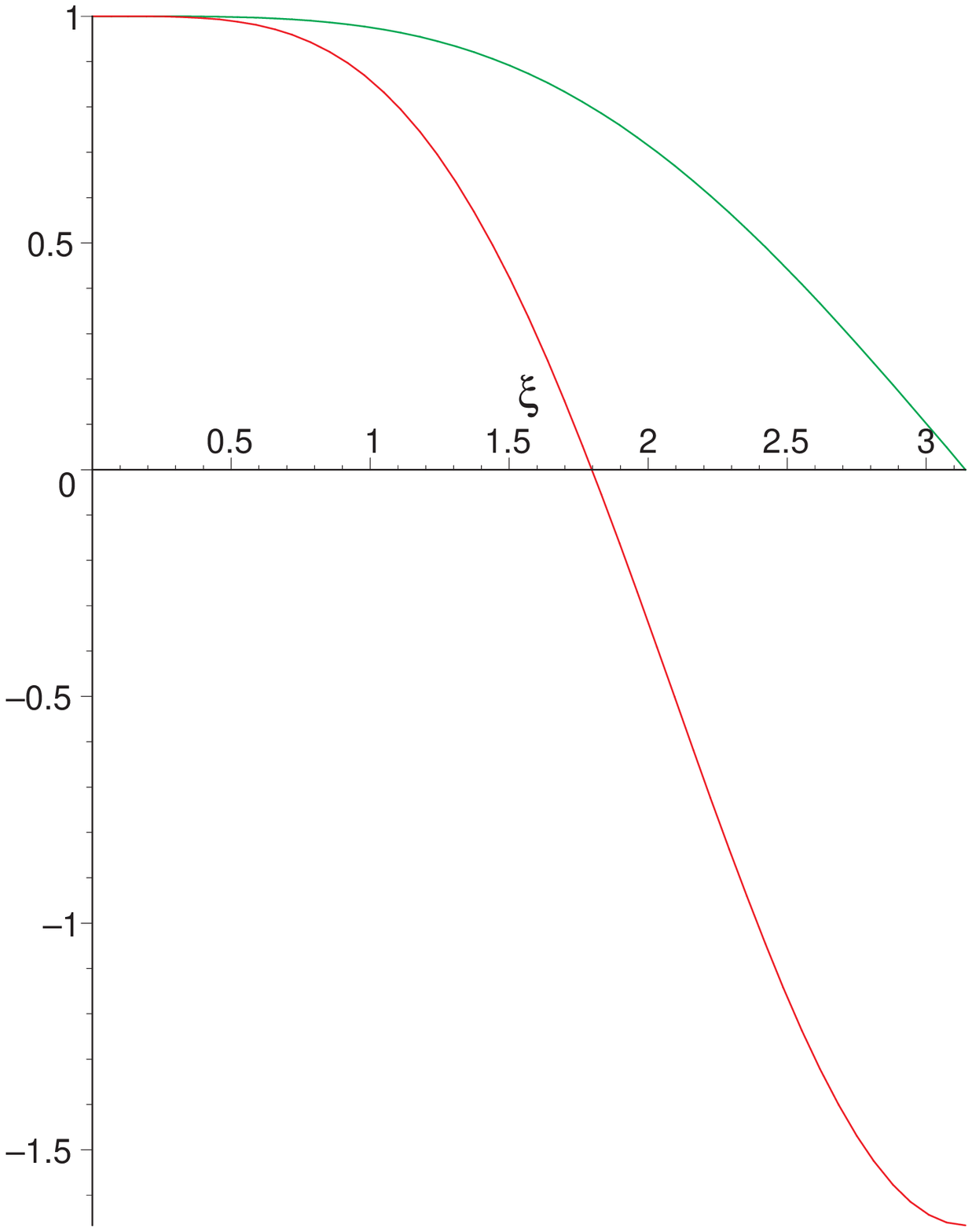}
\end{center}
\caption{The phase (top) and group (bottom) velocities for the second
  (left) and fourth (right) order standard approximation of the
  advective equation (\ref{Eq:advective}) (red) and the wave equation
  (\ref{Eq:wave}) 
  (green).}
\label{Fig:phasegroup}
\end{figure}

Eq.~(\ref{Eq:advective}) arises in the full reduction to first
order of $\phi_{tt} = \phi_{xx}$, while (\ref{Eq:wave}) represents its
reduction in time.  If we denote by $\lambda(\xi)$ an eigenvalue of
the discrete symbol, the corresponding phase and group velocities are
given by
\begin{eqnarray*}
v_{\rm p} &=& i\frac{\lambda(\xi)}{\omega}\,,\\
v_{\rm g} &=& i\frac{d}{d\omega} \lambda(\xi)\,,
\end{eqnarray*}
where $\xi=\omega h$.  In the following table we compute the numerical
phase velocities, $v_{\rm p}$, group velocities, $v_{\rm g}$, the
Courant limits (C.l.), the frequencies of undamped modes (u.m.) and of
the first unstable mode (f.u.m.) for the two systems.  The numerical
phase and group velocities are plotted in Figure \ref{Fig:phasegroup}
as a function of $\xi$.

In the table we used $\Delta^2 = 1 + \frac{1}{3} \sin^2
\frac{\xi}{2}$.  The exact continuum phase and group velocity is 1.
The Taylor expansion of the numerical velocities gives an idea of the
magnitude of the error, provided that enough grid-points per wave
length are used.  The table shows that in the second order accurate
case the phase error for the wave equation is 4 times smaller than for
the advective equation, and that this improvement in accuracy is even
stronger for the fourth order accurate discretization.

Furthermore, the standard discretizations of fully first order
hyperbolic systems have numerical phase velocities that vanish at the
highest frequencies and numerical group velocities with the opposite
sign to the continuum one.  In numerical relativity simulations
involving black holes which make use of the excision technique to
handle the singularity one can expect to see numerical high frequency
solutions escaping from the black hole, if a first order formulation
combined with the standard discretization is used, unless
artificial dissipation is added to the scheme.

Finally, whereas for (\ref{Eq:advective}) the transition from
second order accuracy to fourth order implies the reduction of the
Courant limit by a factor of $1.372$, for the second order in space
system (\ref{Eq:wave}), this transition requires a Courant
limit $2/\sqrt{3} \approx 1.155$ times smaller.  This indicates that
there is an even higher gain in going to fourth order accuracy for second
order in space formulations.

\ifthenelse{\boolean{RevTeX}}
{
  \begin{widetext}
}
{
}

\begin{center}
\begin{tabular}[b]{|c|c|c|c|c|}
\hline
& \multicolumn{2}{|c|}{2nd order accurate} & \multicolumn{2}{|c|}{4th
  order accurate}\\
\hline
& advective & wave & advective &
wave\\
\hline
$v_{\rm p}$ & $\begin{array}{c}\frac{\sin\xi}{\xi} \approx \\1-
  \frac{\xi^2}{6} + O(\xi^4)\end{array}$ &
$\begin{array}{c}\frac{2}{\xi}\sin\frac{\xi}{2} \approx
  \\1-\frac{\xi^2}{24} + O(\xi^4)\end{array}$ &
$\begin{array}{c}\frac{\sin\xi}{\xi}\left( 1 + \frac{2}{3} \sin^2
  \frac{\xi}{2}\right) \\ \approx 1 - \frac{\xi^4}{30} +
  O(\xi^6)\end{array}$ &
$\begin{array}{c}\frac{2}{\xi}\sin\frac{\xi}{2} \Delta \approx \\1 -
  \frac{\xi^4}{180} +  O(\xi^6)\end{array}$\\
\hline
$v_{\rm g}$ & $\begin{array}{c}\cos\xi \approx \\1- \frac{\xi^2}{2} +
  O(\xi^4)\end{array}$ & $\begin{array}{c}\cos\frac{\xi}{2} \approx
  \\1-\frac{\xi^2}{8} + O(\xi^4)\end{array}$ &
$\begin{array}{c}1-\frac{8}{3}\sin^4\frac{\xi}{2} \approx \\1 -
  \frac{\xi^4}{6} + O(\xi^6)\end{array}$ & $\begin{array}{c}
  \cos\frac{\xi}{2} \left(1+
  \frac{2}{3}\sin^2\frac{\xi}{2} \right)/\Delta \\\approx 1 -
  \frac{\xi^4}{36} + O(\xi^6)\end{array}$\\
\hline
C.l. & $\alpha_0$ & $\alpha_0/2$ &
$\alpha_0/1.372$ & $\frac{\sqrt{3}}{4}\alpha_0 \approx
 \alpha_0/2.309$\\
\hline
u.m. & $0,\pi$ & $0$ & $0,\pi$ & $0$\\
\hline
f.u.m.& $\pm \frac{\pi}{2} \approx \pm 1.571$
& $\pi$ &
$\begin{array}{c}\pm 2\arctan\left(\frac{6^{1/4}}{\sqrt{4-\sqrt{6}}}\right)\\
  \approx \pm 1.797\end{array}$ & $\pi$\\ 
\hline
\end{tabular}
\end{center}

\ifthenelse{\boolean{RevTeX}}
{
  \end{widetext}
}
{
}

\section{Discrete constraint propagation}
\label{Sec:ConstrProp}

When simulating systems such as Maxwell's or Einstein's equations, one
has to take into account that the data has to satisfy initial data
constraints.  The evolution equations guarantee that if these
constraints are satisfied initially, then they will be satisfied at
later times.  In this appendix we show that even in the constant
coefficient case, when using standard discretizations of second order
in space systems, the discrete constraints do not propagate exactly.
Initial data which satisfy the discrete constraints do not lead to constraint
satisfying solutions.

As an example, we consider the ADM equations
(\ref{Eq:ADMdens1})--(\ref{Eq:ADMdens2}) with constraints
\[
C \equiv  \frac{1}{2} ( \p_i \p_j \gamma_{ij} - \p_i \p_i
\tau) = 0\,,\qquad
C_i \equiv  \p_j K_{ij} - \p_i K = 0\,.
\]
For simplicity we confine ourselves to solutions which depend only on
one space coordinate.  The discretized constraints are
\begin{eqnarray*}
&&C \equiv -\frac{1}{2} D_+D_- \gamma_{AA} = 0\,, \qquad
C_1 \equiv - D_0 K_{AA} = 0\,,\\
&&C_A \equiv D_0 K_{1A} =0\,,
\end{eqnarray*}
where $A = 2,3$.

The time derivative of the first constraint cannot be expressed in
terms of finite difference combinations of the constraints
\[
\frac{d}{dt} C = D_+D_- K_{AA} \neq -D_0 C_1\,.
\]
This is to be contrasted with the fact that in the constant
coefficient case, the discrete constraints of a first order reduction
would propagate as in the continuum, with partial derivatives replaced
by $D_0$ operators. Furthermore, this issue would not be present if
one used $D_0^2$ to approximate the second derivatives.


\end{document}